\documentclass[preprint,12pt]{elsarticle}

\usepackage{amssymb}
\usepackage{amsmath}

\usepackage[utf8]{inputenc}
\usepackage[T1]{fontenc}

\journal{Physica B: Condensed Matter }

\begin{document}

\begin{frontmatter}

\title{The resistivity of rare earth impurities diluted in Lanthanum (Part I)\tnoteref{t1}}
\tnotetext[t1]{This work was partially funded under the project PIP-11220170100021CO CONICET.}

\author[1]{V P Ramunni\corref{cor1} \fnref{fn1}}
\ead{vivianaramunni@integra.cnea.gob.ar}

\author[2]{A L de Oliveira}

\author[3]{A Troper}

\cortext[cor1]{Corresponding author}
\fntext[fn1]{+54(11) 6772 9278}

\affiliation[1]{organization={Instituto de Nanociencia
	        y Nanotecnolog\'{i}a CNEA - CONICET, 
	        Nodo Constituyentes},
            addressline={Av. Gral. Paz 1499}, 
            city={San Mart\'{\i}n},
            postcode={B1650KNA}, 
            state={Buenos Aires},
            country={Argentina}}
            
\affiliation[2]{organization={Instituto Federal de 
		Educa\c{c}\~{a}o, Ci\^{e}ncia e Tecnologia 
		do Rio de Janeiro},
            addressline={Rua Coronel Delio Menezes Porto 1045}, 
            city={Nil\'opolis},
            postcode={26530-060}, 
            state={RJ},
            country={Brazil}}
            
\affiliation[3]{organization={Centro Brasileiro de Pesquisas 
		F\'{i}sicas},
            addressline={Rua Dr. Xavier Sigaud 150}, 
            city={Rio de Janeiro},
            postcode={22290-180}, 
            state={RJ},
            country={Brazil}}        

\begin{abstract}
We study the temperature-independent resistivity of 
rare-earth magnetic impurities (Gd, Tb, Dy) and the 
non-magnetic impurity (Lu) diluted in double hexagonal 
close-packed (dhcp) lanthanum. We model the system as 
a two-band system, where s-electrons perform conduction 
while d-electrons entirely screen the charge differences induced by impurities. Using the T-matrix formalism derived from the Dyson equation, we obtain an expression for resistivity. Since electronic properties are highly sensitive to band structure, we examine two types: a simplified ``parabolic'' band structure and a more realistic one obtained through first-principles 
calculations using VASP. Our results indicate that the 
exchange parameters, represented as cross products, 
significantly influence the magnitude of the spin 
resistivity term. Furthermore, we find that the role of 
band structure in resonant scattering and the formation 
of virtual bound states depends on the specific band 
structure employed. Additionally, our study addresses the effects of translational symmetry breaking and the excess charge introduced by rare-earth impurities on resistivity.
\end{abstract}

\begin{highlights}
\item Resistivity of rare earth impurities diluted in Lanthanum
\item Two types of density of states (DOS) were considered
\item Influence of density of states on resistivity
\item Drastic change of resistivity with the DOS model
\item Changes of standard result of De Gennes and Friedel with DOS.
\end{highlights}

\begin{keyword}
Residual and spin disorder resistivity \sep Lanthanum host 
\sep Rare-earth impurities \sep Period and volume effects 
\sep Ab-initio calculations \sep Density of states models. 
\end{keyword}

\end{frontmatter}
\section{Introduction}
\label{S0}

Recent studies on the electronic properties of lanthanum alloys doped with gadolinium (Gd), dysprosium (Dy), terbium (Tb), and lutetium (Lu) have highlighted significant advances in their magnetocaloric and superconducting behaviors. Each of these heavy lanthanides imparts unique magnetic and electronic characteristics to lanthanum-based alloys, largely due to the unfilled 4f electron shells of the rare earth elements. These properties make lanthanum alloys highly valuable in various technological applications, particularly in areas requiring strong magnetic properties or specialized conductivity.

In addition to these magnetic effects, lanthanum compounds have shown remarkable progress in the field of superconductivity. Notably, lanthanum superhydrides, such as $LaH_{10}$, which are synthesized under high hydrogen content and extreme pressures, have demonstrated superconducting states around 250 K. This discovery has spurred further exploration into the potential of lanthanum-based materials for high-temperature superconductivity  \cite{PRL,GUO22}. 

The production of rare earth elements is fundamental to the advancement of many modern technologies. High-growth industrial sectors, including electric vehicles, wind turbines, and data centers, rely heavily on rare-earth-based high-performance magnets \cite{SIMS22}. Recent research suggests that lanthanum could help diversify the rare earth mining economy and enhance supply chain stability for all rare earth elements, particularly in applications involving aluminum-based alloys. Using lanthanum in manufacturing offers multiple advantages \cite{SIMS22}.

Furthermore, rare-earth monopnictides have attracted significant interest due to their distinctive electronic and topological properties, making them promising candidates for device applications. Recently, Lin et al. examined rock-salt structured lanthanum monopnictides LaX (X = P, As) through density functional theory (DFT) simulations, as discussed in Ref. \cite{LIN21}. Thermoelectric materials, which can directly convert heat into electricity, have a range of existing and potential applications in the energy sector and industrial power generation \cite{LIN21}.

Although in past decades, Lanthanum has been extensively studied due to its superconducting properties in alloys like La-Y and La-Lu \cite{ANDERSON58}, less studied, incorporating Gd, Tb, and Dy into such lanthanum structures may further stabilize the superconducting phase due to the increased electron density and magnetic properties. Related to the electronic state modifications, adding Gd, Tb, and Dy to lanthanum affects its density of states and can result in changes to magnetic ordering and electronic band structure, influencing conductivity and magnetization properties. These effects are complex, as they vary depending on pressure, temperature, and alloying concentrations, but they open pathways for custom-engineered electronic and magnetic properties in lanthanum alloys. 
Concerning electrical conductivity and resistivity, the rare-earth impurities, when alloyed with lanthanum, could influence strongly these properties which are essential for designing components in electronic devices and developing new superconductors materials and energy storage systems. These diluted alloys could provide new insights into improving the performance, efficiency, or stability of materials in these applications.
This study of resistivity aims to investigate the resistivity properties of gadolinium (Gd), dysprosium (Dy), lutetium (Lu), and terbium (Tb) when diluted in lanthanum. By systematically examining these rare-earth elements, we aim to contribute to a better understanding of resistivity behaviors in lanthanide-based systems and provide data that could support future applications in advanced materials.
With this idea in mind, as a starting point for our future investigations, we have extended the expression for the temperature-independent resistivity, $R$, associated with rare-earth impurities (Gd, Tb, Dy, and Lu) embedded in a transition metal-like host, building on the previous work by Troper and Gomes~\cite{Atroper}. Our model describes a two-band paramagnetic metallic system consisting of s- and d-bands, which we simplify by assuming these bands are non-hybridized and coupled through the usual exchange interactions, $J^{(s)}$ and $J^{(d)}$, to the local 4f-spin of the magnetic rare-earth impurity. We have assumed that conduction is entirely due to s-electrons, while d-electrons screen the charge differences induced by the impurities.
The degeneracy of the d-band can be incorporated straightforwardly by employing the standard approximation of five identical sub-bands. In addition to the model presented in Ref.~\cite{Atroper}, we consider two significant charge effects introduced by the impurities first, the charge difference between the impurity and host and second, the translational symmetry breaking, which leads to a non-local charge potential \cite{Zeller}. Furthermore, we adopt a phenomenological approach to account for the volume difference between the impurity and the host~\cite{Daniel}.
For the ``magnetic'' rare-earth impurities, we include a spin-dependent potential addressed using first-order perturbation theory, as is standard practice \cite{Blackman}.
An essential characteristic of rare-earth elements embedded in transition metal hosts, such as lanthanum, is the charge difference between the host and the impurity, which is typically trivalent. The simultaneous presence of spin and charge potential scattering is a key signature of the behavior of such impurities in transition metal hosts.
In this way, we investigate the role of band structure in resonant scattering, particularly in the context of virtual bound states, by employing a realistic band structure. By varying the position of the Fermi level within the band structure to explore the vicinity of the band edge, we can identify regions with narrow peaks and other features, thereby allowing for a thorough discussion of these details in our calculations.
In summary, we examine the dependence of temperature-independent resistivity on the band structure of the 5d host. Our study focuses on diluted lanthanum alloys containing rare-earth impurities such as Gd, Tb, Dy, and Lu. To this end, we utilize two models for the band structure: one based on a ``parabolic'' (vastly used in the literature) band and another, more realistic, derived from first-principles calculations using the VASP code~\cite{VASP1, VASP2, VASP3, VASP4}.
In deriving the expression for temperature-independent resistivity, we assume that the magnetic impurity, positioned at the origin, contributes to two scattering sources. The first is a spin-independent charge potential, $R_0$, which results from the charge difference introduced by the impurity. The second source is the potential scattering through d-d interactions, represented by $V_{dd}$, calculated using an extended Friedel sum rule. Additionally, we account for impurity-induced mixing, denoted as $V_{sd}$. Consequently, the expression for $R$ in equation~(\ref{Eq0}) can be written as follows:
\begin{equation}
R(\epsilon) = R_0 (\epsilon) + \pi Ac\rho _s(\epsilon)S_0 (S_0+1)\left(J^{(s)}\right)^2 \times \left[ 1+\theta (\epsilon, \vert V_{sd}\vert ^2) \right].
\label{Eq0} 
\end{equation}
In equation~(\ref{Eq0}), $A$ is a constant that depends on the characteristics of the s-Fermi surface of the host, while $c$ denotes the concentration of impurities diluted in the host. The spin scattering term arises from conduction electrons that couple to the impurity spin, $S_0$, via k-independent exchange interactions $J^{(s)}$ and $J^{(d)}$. The function $\theta (\epsilon;\vert V_{sd}\vert ^2)$ describes the combined effects of the charge potential and spin-dependent scattering, which influence the deviations from the De~Gennes Friedel resistivity, given by
\begin{equation}
R_{DG-F}=\pi Ac(J^{(s)})^2S_0(S_0+1).
\label{RDGF0} 
\end{equation}
First, we discuss the role of band structure in resonant scattering, particularly in relation to virtual bound states, by adopting two types of density of states: a ``parabolic'' model and a more realistic band structure calculated using VASP. We assume that the rare-earth impurity behaves as a ``spherically symmetric'' entity; therefore, we disregard effects such as skew scattering (related to spin-orbit coupling) and crystal field splitting. Consequently, our calculations most apply to the Gd impurity in transition metal-like hosts, where the total spin is $S_0(Gd)=7/2$.
However, we can also consider impurities as Tb and Dy, which have non-zero orbital angular momentum ($L \neq 0$). In these cases, we replace $S_0$ with the effective spin $(g_j-1)J_0= S_0^{eff}$, where $(g_j-1)$ is the De~Gennes factor that represents the projection of the spin $S_0$ onto the total angular momentum $J_0$ of Tb and Dy impurities. Thus, we have  $S_0(Gd)=7/2$, $S^{eff}_0(Tb)=6/2$, and $S^{eff}_0(Dy)=5/2$. It is important to note that both $S_0$ and $S^{eff}_0$ are good quantum numbers, as they arise from Hund's rules.
Our results confirm a strong dependence of resistivity on the band structure and, consequently, on the shape of the density of states employed. This conclusion is supported by the observation that:

\begin{enumerate}[1.]
\item Our expression for temperature-independent resistivity, formulated in terms of the hybridization charge potential $\vert  V_{sd}\vert ^2$, exhibits significant variations depending on whether we use a parabolic or a realistic model for the density of states (DOS).
\item We investigate the exact form of the resistivity, $R=R(\epsilon,\vert V_{sd}\vert ^2)$, with the approximate case, $R=\vert V_{sd}\vert ^2=\vert V_{sd}\vert ^2 R(\epsilon)$, up to first order in $\vert V_{sd}\vert ^2$. We observe that for the DOS-Moriya model, the interval of $\vert V_{sd}\vert ^2$ values in which $R\simeq R_{app}$ is almost the same as compared to that of the realistic DOS. 
\item The significant impact of the DOS on the temperature-independent resistivity $R$ has motivated a systematic study of transition metal hosts belonging to the 5d series, including Hf, Ta, W, Re, Ir, Os, Pt, and Au. The results of this study will also be shared in future work.
\end{enumerate}
The rest of the manuscript is structured as follows: Section~\ref{S2} introduces the two-band model used to derive an expression for temperature-independent resistivity, which is summarized in Section~\ref{S3}. Section~\ref{S4} outlines the computational procedure and the convergence tests performed to determine the energy and k-mesh parameters from density functional theory (DFT) calculations using VASP, applicable to both the primitive and supercell structures. We then present the results of band structure calculations based on the previously relaxed structure of double hexagonal close-packed (dhcp) lanthanum.
Section~\ref{S5},  details the critical potential and the calculated resistivity of the diluted impurities (Gd, Tb, Dy, and Lu) in La. We employ both the ``parabolic'' model and the model based on DFT calculations for the s- and d-density of states of lanthanum. We adopt the approximation of homothetic bands, $\epsilon _{dk}=\alpha \epsilon _{sk}$, where $\alpha $ is a phenomenological parameter representing the ratio of the effective masses of s(p)- and d-electrons. While $\epsilon _{\lambda k}$ denotes the dispersion relation of the $\lambda$-electrons (with $\lambda = $s, d). Finally, Section~\ref{S6} presents our conclusions. We also include two Appendices to provide a comprehensive description of our system.
\section{The two-band Hamiltonian model}
\label{S2}

In this section, we introduce the two-band model that will be used to derive the expression for temperature-independent resistivity. For simplicity, we adopt a two-band model for the host material, assuming that the s- and d-bands are not hybridized. Since the s-electrons are primarily responsible for conduction, we focus on deriving a Dyson-like equation for the one-electron s-s perturbed propagators in terms of the host metal.

This approach is commonly applied in the study of resistivity and electronic transport in metals, where separate conduction bands (s and d) are treated independently, particularly when the s-band dominates the electrical conductivity. By simplifying the model in this way, we aim to capture the essential physics while avoiding unnecessary complexity, allowing for clearer insights into the temperature-independent resistivity of the material. In our model, the full Hamiltonian is expressed as detailed in Ref.~\cite{Atroper}.
In our model, the full Hamiltonian is expressed as detailed in Ref. \cite{Atroper}.

\begin{equation} 
H = H_{0}^{(s,d)}+H_{I}^{q}+H_{I}^{\tau }+H_{I}^{\vec{S}_{0}}, 
\label{HT}
\end{equation} 
This Hamiltonian intends to be model for rare earths diluted in transition
metal like host or intermetallic compounds. 
In equation~(\ref{HT}), the pure host is described by the following Hamiltonian in the Wannier notation:  
\begin{equation}  
H_{0}^{(s,d)}=\sum_{i,j,\sigma }t_{ij}^{(s)}c_{i\sigma }^{\dagger }c_{j\sigma  
}+\sum_{i,j,\sigma }t_{ij}^{(d)}d_{i\sigma }^{\dagger }d_{j\sigma }, 
\end{equation}  
where $c_{i\sigma }$,$c_{i\sigma }^{\dagger }$ and $d_{i\sigma }$,$d_{i\sigma }^{\dagger} $ stand for the creation/aniquilation operators of an electron with polarization $\sigma$ on the site $i$ for the s and d band respectively, 
with $t_{ij}^{(\lambda )}$ representing the transfer integral of $\lambda $-electrons between the host's sites $i$ and $j$. 
A magnetic impurity located at the origin introduces two sources of scattering, first the spin-independent potential $V_{dd}$, which arises from the charge difference caused by the impurity, and second the potential scattering due to d-d interactions. Additionally, impurity-induced s-d mixing contributes to the scattering processes as follows:
\begin{equation}  
H_{I}^{q}=V_{dd}\sum_{\sigma }n_{0\sigma }^{(d)}+\sum_{\sigma }\left\{  
V_{sd}c_{0\sigma }^{\dagger }d_{0\sigma }+V_{ds}d_{0\sigma }^{\dagger  
}c_{0\sigma }\right\}; 
\end{equation}
with $n_{0\sigma }^{(d)}$ the number of electrons at the impurity site in the d-band.
Thus, in $H_{I}^{q}$ inter s-d band effects are described through a one-body hybridization term $V_{sd}$, diverging from the conventional many-body approach, and we treat the hybridization within the independent approximation of the wave vector $\vec{k}$.
The second source of scattering introduced by the impurity is the Spin scattering term. The conduction electrons are coupled to the impurity spin through $k,k^{\prime}$-independent exchange interactions. The corresponding Hamiltonian is,
\begin{equation}  
H_{I}^{\vec{S } _{0}}=J^{(s)}\vec{\sigma }^{(s)}\cdot \vec{S}_{0}+J^{(d)}\vec{\sigma } ^{(d)}\cdot \vec{S}_{0},
\label{ISPIN}  
\end{equation}  
where the constants $J^{(s,d)}$ represent the exchange integrals for the s- and d-electrons, respectively. 
The parameter $\vec{S}_{0}$, denotes the local impurity spin with components $\sigma _{i}$. 
\begin{equation}  
\sigma ^{z}_{i}=a^{\dagger }_{i \uparrow }a_{i \uparrow }-a^{\dagger }_{i \downarrow }a_{i \downarrow }, \,\,
\sigma ^{+}_{i}=a^{\dagger }_{i \uparrow }a_{i \downarrow } , \,\,    
\sigma ^{-}_{i}=a^{\dagger }_{i \downarrow }a_{i \uparrow }.  
\label{S01}
\end{equation}   
The $a_i ^{(\lambda)}$'s in the above expression stand for the $c_{i}$ or $d_{i}$ operators, corresponding to s- and d-states, respectively ($\lambda=$s or d).
We have extended the Hamiltonian presented in Ref.~\cite{Atroper} by incorporating two primary charge effects introduced by the impurities, namely  the breaking of translational symmetry,
\begin{equation}  
H_{I}^{\tau }=\tau \sum_{m\neq 0,\sigma }\left\{ t_{0m}^{(d)}d_{0\sigma  
}^{\dagger }d_{m\sigma }+t_{m0}^{(d)}d_{m\sigma }^{\dagger }d_{0\sigma  
}\right\},
\end{equation}
and the volume difference between the impurity ($\Omega ^{I}$) and host ($\Omega $), which is incorporated into the effective charge difference as,
\begin{equation}
\Delta Z^{\prime }=\Delta Z-\frac{\Omega ^{I}-\Omega}{\Omega}=\Delta Z-\frac{\delta v}{\Omega},
\label{Eq8} 
\end{equation}
being $\Delta Z$ calculated through,
\begin{equation}
\Delta Z = \frac{2}{\pi} \tan ^{-1}\frac{\pi \vert V_{dd}(\omega )\vert  \rho _d(\omega )}{1-\vert V_{dd}(\omega)\vert g_{00}^{dd(R)}(\omega)}.
\label{Eq9}
\end{equation}
In equation~(\ref{Eq9}), $\Delta Z$ is the real charge difference and $ \Delta Z^{\prime}$ is the effective charge difference between impurity and host elements. 
In the absence of spin scattering the solution, in terms of the perfect lattice propagators $P^{\lambda \lambda}_{ij}$ (with $\lambda=s,d$) is given by, 
\begin{equation} 
G_{ij}^{ss}(\omega)=P_{ij}^{ss}(\omega)+2\pi P_{i0}^{ss}(\omega)T^{ss}(\omega)P_{0j}^{ss}(\omega)
\label{G1}
\end{equation}
and
\begin{equation}
G_{ij}^{ds}(\omega)=2\pi P_{i0}^{dd}(\omega)T^{ds}(\omega)P_{0j}^{ss}(\omega),
\label{G2}
\end{equation}
with
\begin{equation}
P^{\lambda \lambda}_{ij}=\frac{1}{2\pi}\sum_k \frac{e ^{ik(R_i-R_j)}}{\omega -\epsilon ^{\lambda}_k},
\label{Cauchy}
\end{equation}
and
\begin{equation}
\omega ^{\lambda}_k=\sum_{\delta} T^{\lambda}_{0\delta} e^{ikR_{\delta}},
\label{disp}
\end{equation}
being the Cauchy's principal part $P$ and the energy band structure $\omega _k ^{\lambda}$ in equations~(\ref{Cauchy}) and (\ref{disp}), respectively.
\noindent The $T(\omega)$ scattering matrices in equations (\ref{G1}) and (\ref{G2}) are defined as follows:
\begin{equation}  
T^{ss}(\omega)=\frac{\vert V_{sd}\vert ^{2}P_{00}^{dd}(\omega)}{1-\left[V_{dd} + \vert V_{sd}\vert ^{2} P_{00}^{ss}(\omega)\right]P_{00}^{dd}(\omega)}  
\label{G3}
\end{equation}  
and,
\begin{equation}  
T^{ds}(\omega)=\frac{V_{ds}}{1-\left[V_{dd} + \vert V_{sd}\vert ^{2} P_{00}^{ss}(\omega)\right]P_{00}^{dd}(\omega)}.
\label{G4}  
\end{equation}  
Potential scattering (which may be rather strong) arises from the difference of valence between the host and the trivalent rare earth. 
The second set of equations for the d-d and mixed s-d interactions is given by:
\begin{equation}
G_{ij}^{dd}(\omega)=P_{ij}^{dd}(\omega)+2\pi P_{i0}^{dd}(\omega)T^{dd}(\omega)P_{0j}^{dd}(\omega) 
\label{G5}
\end{equation} 
and
\begin{equation}
G_{ij}^{sd}(\omega)=2\pi P_{i0}^{ss}(\omega)T^{sd}(\omega)P_{0j}^{dd}(\omega).
\label{G6}
\end{equation} 
The $T$ matrices in equations (\ref{G5}) and (\ref{G6}) are defined as,
\begin{equation}  
T^{dd}(\omega)=\frac{V_{dd}+\vert V_{sd}\vert   ^{2}P_{00}^{ss}(\omega)}{1-\left[V_{dd} + \vert V_{sd}\vert   ^{2} P_{00}^{ss}(\omega)\right]P_{00}^{dd}(\omega)}
\label{G7}
\end{equation}  
and
\begin{equation}  
T^{sd}(\omega)=\frac{V_{sd}}{1-\left[V_{dd} + \vert V_{sd}\vert   ^{2} P_{00}^{ss}(\omega)\right]P_{00}^{dd}(\omega)}.
\label{G8}  
\end{equation}
As previously demonstrated in Ref.~\cite{Atroper}, the problem of an impurity diluted in a transition metal lattice, in the absence of magnetic order $<S^{z}_0>$, is entirely solved in terms of $V_{dd}$ and $V_{sd}$, along with the (s,d)-band structures $\epsilon _{sk}$, $\epsilon _{dk}$, respectively.
At zero order in $\tau $, the term $H^{\tau}_I$ in the total Hamiltonian renormalizes the local charge potential $V_{dd}$ of the impurity located at the origin, resulting in the energy-dependent potential $V_{dd}(\omega)$. Consequently, the propagators are redefined in terms of the effective non-local charge potentials $V_{dd}(\omega)$ and $V_{sd}$ as follows:
\begin{equation}  
\tilde{G}_{ij}^{sd}(\omega)=P_{ij}^{ss}(\omega)+P_{i0}^{ss}(\omega)V_{ss}\tilde{G}_{0j}^{ds}(\omega)  \label{Gijss}
\end{equation}
\noindent and for the mixed term,
\begin{equation}  
\begin{array}{lll}  
\tilde{G}_{ij}^{ds}(\omega) & = & P_{i0}^{dd}(\omega)V_{dd}\tilde{G}_{0j}^{ds}(\omega)+P_{i0}^{dd}(\omega)V_{ds}\tilde{G}_{0j}^{ss}(\omega) \\ 
& + & \tau P_{i0}^{dd}(\omega)\left\{ \sum_{m\neq 0,\sigma }t_{0m}^{(d)}\tilde{G}_{mj}^{ds}(\omega)\right\} 
+\tau \left\{ \sum_{m\neq 0,\sigma }P_{im}^{dd}(\omega)t_{m0}^{(d)} \right\} \tilde{G}_{0j}^{ds}(\omega),  \label{Gtildeds}
\end{array}  
\end{equation}  
\noindent with
\begin{eqnarray}
\sum_{m \neq 0, \sigma } t_{0m}^{(d)}\tilde{G}_{mj}^{ds}(\omega) & =
 & \frac{1}{(\tau +1)}\left\{(\omega-\tilde{\epsilon }_{d}-U_{ds})\tilde{G}_{0j}^{sd}(\omega)-\delta _{0j}\right\}, \label{Zefsd1} \\  
\sum_{m \neq 0, \sigma } P_{im}^{dd}(\omega)t_{m0}^{(d)} & = & (\omega -\tilde{\epsilon }_{d})P_{i0}^{dd}(\omega)-\delta _{i0}  \label{Zefsd2} 
\end{eqnarray} 
\noindent and
\begin{equation}  
U_ {ds}(\omega)=V_{dd}+P_{00}^{ss}(\omega)\vert V_{ds}\vert   ^2.  \label{vefsd}
\end{equation}  
The new non-local charge potentials are then,
\begin{equation}
V_{ss}(\omega)=\frac{\vert V_{sd}\vert ^2 P_{00}^{dd}(\omega)}{\alpha ^2 - \left[ V_{dd}+(\alpha ^2-1)(\omega-\epsilon _{c}^{d})\right]P_{00}^{dd}(\omega)} ,
\end{equation}
\begin{equation}
V_{dd}(\omega)=V_{dd}+(\alpha ^2-1)(\omega -\epsilon _{c}^{d})\,,
\label{Veff1}
\end{equation} 
which accounts for the breaking of translational invariance in the s- and d-bands. 
Here, $\tau =\alpha -1$ serves as a proportionality factor that facilitates the renormalization of the impurity-host hopping in relation to the host-host hopping. 
In terms of the new effective scattering matrices $T_{ss}$ and $T_{ds}$, the Green functions are expressed as,
\begin{equation} 
\tilde{G}_{ij}^{ss}(\omega)=P_{ij}^{ss}(\omega)+P_{i0}^{ss}(\omega)T_{ss}(\omega)P_{0j}^{ss}(\omega) +\theta (\tau ), 
\label{GTss}
\end{equation}  
\begin{equation} 
\tilde{G}_{ij}^{ds}(\omega)=P_{i0}^{dd}(\omega)T_{ds}(\omega)P_{0j}^{ss}(\omega) +\theta (\tau ),
\label{GTds}
\end{equation} 
with the scattering matrices defined by,
\begin{equation}
T_{ss}(\omega)=\frac{\vert V_{sd}\vert   ^2 P_{00}^{dd}(\omega)}{ \alpha ^2 -\left[V_{dd}(\omega)+ \vert V_{sd}\vert   ^2 P_{00}^{ss}(\omega)\right]P_{00}^{dd}(\omega)},
\end{equation} 
\begin{equation}
T_{ds}(\omega)=\frac{V_{ds}}{ \alpha ^2-\left[V_{dd}(\omega)+ \vert V_{sd}\vert   ^2 P_{00}^{ss}(\omega)\right]P_{00}^{dd}(\omega)}. 
\end{equation} 
The full set of Dyson's equations in the absence of spin scattering is expressed as, 
\begin{equation}  
\begin{array}{lll}  
\tilde{G}_{ij}^{dd}(\omega) & = & P_{ij}^{dd}(\omega) + P_{i0}^{dd}(\omega)V_{dd}\tilde{G}_{0j}^{dd}(\omega)+P_{i0}^{dd}(\omega)V_{ds}\tilde{G}_{0j}^{sd}(\omega) \label{22} \\  
& + & \tau P_{i0}^{dd}(\omega)\left\{ \sum_{m\neq 0,\sigma }t_{0m}^{(d)}\tilde{G}_{mj}^{dd}(\omega)\right\} +\tau \left\{ \sum_{m\neq 0,\sigma }P_{im}^{dd}(\omega)t_{m0}^{(d)} \right\} \tilde{G}_{0j}^{dd}(\omega)  
\end{array}  
\end{equation}  
with 
\begin{equation}
\tilde{G}_{ij}^{sd}(\omega)=P_{ij}^{ss}(\omega)V_{sd}\tilde{G}_{0j}^{dd}(\omega). 
\label{44} 
\end{equation} 
Replacing equation~(\ref{44}) in equation~(\ref{22}), for $i=0$ we obtain
\begin{equation}  
\begin{array}{lll}  
\tilde{G}_{ij}^{dd}(\omega) & = & P_{ij}^{dd}(\omega) + P_{i0}^{dd}(\omega) \left\{ V_{dd} + \vert V_{ds}\vert   ^2 P_{00}^{ss}(\omega) \right\} \tilde{G}_{0j}^{dd}(\omega) \\ 
& + & \tau P_{i0}^{dd}(\omega)\left\{ \sum_{m\neq 0,\sigma }t_{0m}^{(d)}\tilde{G}_{mj}^{dd}(\omega)\right\} +\tau \left\{ \sum_{m\neq 0,\sigma }P_{im}^{dd}(\omega)t_{m0}^{(d)} \right\} \tilde{G}_{0j}^{dd}(\omega).
\end{array} 
\label{Gdd-FRIEDEL} 
\end{equation}  
Using Zeller's relations for the d-band~\cite{Zeller},
\begin{eqnarray}  
\sum_{m\neq 0,\sigma }t_{0m}^{(d)}G_{mj}^{dd}(\omega) &=&\frac{1}{(\tau +1)}\left\{ (\omega -\tilde{\epsilon } _{s}-U_{dd})G_{0j}^{dd}(\omega)-\delta  
_{0j}\right\} \,, \\  
\sum_{m\neq 0,\sigma }P_{im}^{dd}(\omega)t_{m0}^{(d)} &=&(\omega -\tilde{\epsilon}_{s})P_{i0}^{dd}(\omega)-\delta _{i0}\,,  
\end{eqnarray}  
\noindent where $U_{dd}$ is defined as, 
\begin{equation}
U_{dd}(\omega) = V_{dd} + \vert V_{ds}\vert   ^2 P_{00}^{ss}(\omega).
\end{equation} 
Considering the following approximation,
\begin{equation}  
\left( \frac{\alpha -1}{\alpha }\right) =\frac{\tau }{\tau +1}\sim \tau  
-\tau ^{2}+\cdots , 
\end{equation}  
and keeping only the first-order term in $\tau$, we obtain the complete equation for the spin-independent charge potential up to first order in $\tau $ as,
\begin{equation}
\widetilde{G}_{ij}^{dd}(\omega)=P_{ij}^{dd}(\omega)+P_{i0}^{dd}(\omega) T_{dd}(\omega) P_{0j}^{dd}(\omega)+\theta (\tau )
\label{GTdd}
\end{equation} 
and 
\begin{equation}  
\widetilde{G}_{ij}^{sd}(\omega)=P_{i0}^{ss}(\omega)T_{sd}(\omega)P_{0j}^{dd}(\omega)+\theta (\tau ),  
\label{GTsd}
\end{equation}  
with $T_{dd}$ e $T_{sd}$ representing the effective scattering matrices, 
\begin{equation}
T_{dd}(\omega)=\frac{V_{dd}(\omega)}{\alpha ^2 - V^{\star}_{dd}(\omega)P_{00}^{dd}(\omega)} 
\end{equation} 
and 
\begin{equation}
T_{sd}(\omega)=\frac{\vert V_{sd}\vert   }{\alpha ^2-V^{\star}_{dd}(\omega)P_{00}^{dd}(\omega)} \,,
\end{equation} 
in terms of the new non-local charge potential $V_{dd}^{\star}(\omega) $
\begin{equation}
V_{dd}^{\star}(\omega)=V_{dd}+(\alpha ^2 -1)(\omega-\epsilon _c ^d)+\vert V_{sd}\vert ^2 P_{00}^{ss}(\omega)=V_{dd}(\omega )+\vert V_{sd}\vert ^2 P_{00}^{ss}(\omega),
\label{Veff2}
\end{equation} 
with $V_{dd}(\omega)$ defined in equation (\ref{Veff1}).
At zero order in $\tau$, the term $H_{I}^{q}$ in the Hamiltonian $H = H_{0}^{(s,d)}+H_{I}^{q}+H_{I}^{\tau }+H_{I}^{\vec{S}_{0}}$ in equation (\ref{HT}), simply renormalizes the impurity charge potential $V_{dd}$ at the site $i=j=0$ by $V_{dd}(\omega)$. 

 In the absence of spin scattering, the impurity problem is fully resolved by equations (\ref{GTss}), (\ref{GTds}), (\ref{GTdd}), (\ref{GTsd}) in terms of $V_{dd}$, $V_{sd}$, and the band structure of the host through $\epsilon ^{(s)}_k$ and $\epsilon ^{(d)}_k$. Finally, the integral equation for the complete Hamiltonian is given by:
\begin{equation}
\omega \Gamma _{ij}^{ss}(\omega )=\frac{1}{2\pi}\delta _{ij} + \sum _l T^{(s)}_{il}\Gamma _{lj}^{ss}(\omega )+\delta _{i0}V_{sd} \Gamma _{0j}^{ds}(\omega ) + J^{(s)}\delta _{i0} \theta _{0j}^{ss}(\omega ),  
\label{GSpina}
\end{equation}
and
\begin{equation}
\omega \Gamma _{ij}^{ds}(\omega)=\sum _l T^{(d)}_{ij}\Gamma _{lj}^{ds}(\omega )+\delta _{i0}V_{dd} \Gamma _{0j}^{ds}(\omega ) + \delta _{i0}V_{ds} \Gamma _{0j}^{ss}(\omega ) + J^{(d)}\delta _{i0} \theta _{0j}^{ds}(\omega ). 
\label{GSpinb}
\end{equation}
These two last exact equations of motion, given by (\ref{GSpina}) and (\ref{GSpinb}), fully specify the propagators $\Gamma _{ij}^{ss}$ and  $\Gamma _{ij}^{ds}$ in terms of the  ``spin-flip'' propagators $\theta _{ij}$.
We can now derive approximate equations of motion for the $\theta _{ij}^{ss}$ and $\theta _{ij}^{ds}$ Green's functions. Following the procedure outlined by Blackman and Elliott~\cite{Blackman}, we decouple higher-order propagators in accordance with Nagaoka’s scheme~\cite{NAGAOKA}, yielding:
\begin{eqnarray}
\Gamma ^{ss}(\omega) & = & G^{ss}_{ij} + 4\pi ^2 S_0 (S_0+1) G^{ss}_{i0}J^{(s)}G^{ss}_{00}J^{(s)}G^{ss}_{0j} \\
                     & + & 4\pi ^2 S_0 (S_0+1) G^{sd}_{i0}J^{(d)}G^{ds}_{00}J^{(s)}G^{ss}_{0j}  \nonumber \\
                     & + & 4\pi ^2 S_0 (S_0+1) G^{ss}_{i0}J^{(s)}G^{sd}_{00}J^{(d)}G^{ds}_{0j}  \nonumber \\
                     & + & 4\pi ^2 S_0 (S_0+1) G^{sd}_{i0}J^{(d)}G^{dd}_{00}J^{(d)}G^{ds}_{0j}. \nonumber \\
\label{TssSpin}
\end{eqnarray}

Finally, the complete expression for the $G^{ss}(\omega )$ propagator in terms of the scattering matrix $T(\omega)$ is given by:
\begin{equation} 
\tilde{G} _{ij}^{ss}(\omega)=P_{ij}^{ss}(\omega)+P_{i0}^{ss}(\omega) T(\omega) P_{0j}^{ss}(\omega), 
\label{GSS}
\end{equation} 
where $T$ is defined as follows,  
\begin{eqnarray} 
T(\omega) & = & \frac{\vert V_{sd}\vert   ^2 F_{d}(\omega)}{X(\omega)}  
+ S_0 (S_0 +1)\left\{ \frac{ [J^{(s)}]^2[1-V_{dd}(\omega)F_d(\omega)]^3F_s(\omega)}{X^3(\omega)}      \right. \nonumber \\
& + & \frac{2\vert V_{sd}\vert ^2J^{(s)}J^{(d)} F_s(\omega)[F_d(\omega)]^2[1-V_{dd}(\omega)F_d(\omega)]}{X^3(\omega)} \nonumber \\
& + & \left. \frac{[J^{(d)}]^2 \vert V_{sd}\vert ^2 [F_d(\omega)]^3 }{X^3(\omega)} \right\},
\label{T2b}
\end{eqnarray}
and
\begin{equation}
X (\omega)= 1-V_{dd}(\omega)P^{dd}_{00}(\omega)-\vert V_{sd}\vert   ^2P^{ss}_{00}(\omega)P^{dd}_{00}(\omega),
\label{Xeq}
\end{equation}
with
\begin{equation}
 F_{\lambda }(\omega)=\sum_k \frac{1}{\omega -\epsilon ^{\lambda }_k}=P^{\lambda }_{00}(\omega); \,\,\, \lambda=s,d. 
\end{equation}
The terms $\omega = \epsilon \pm i\delta $; $F^I_s (\epsilon _F)=\pi \rho _s (\epsilon _F)$, $\vert   F_{\lambda}\vert $ and $\delta _{\lambda}$ ($\lambda = s,d$), are associated with the band structure of the host, while $\vert X(\omega) \vert $ and $\vert \eta _{dd}\vert $ relate to the charge potential introduced by the impurity.
In deriving equation~(\ref{GSS}), we focus on formulating a Dyson-like equation for the one-electron s-s perturbed propagators in terms of the host metal propagators. This equation is valid for all orders in perturbation theory and adheres to the accuracy of Nagaoka’s decoupling scheme.

Note that, in our model we neglect the term $H^{Imp}_{Coul.} = \Delta U n^{(d)}_{0\uparrow}n^{(d)}_{0\downarrow}$ where $\Delta U=U_{imp} - U_{host}$, represents the difference in the Coulomb repulsion parameter at the impurity site \cite{Troper78}. For rare-earths impurities like Gd and Lu diluted in 5d metals one expects $U \simeq 0$; so the results including finite $U$ are expected to apply for rare-earths in, say, 3d and 4d metals. 

\section{Theoretical results: A brief Account }
\label{S3}
In the present Section we derive the expression of the resistivity for rare-earth impurities diluted in lanthanum host from the scattering matrix $T(\omega )$ in equation (\ref{T2b}) previously obtained in Section \ref{S2}.
In the absence of magnetic order ($<S^z>=0$) up to first order in $J^2$, the  temperature independent resistivity for a small concentration of impurities $c$ randomly distributed within the host, is expressed as follows~\cite{Atroper}:
\begin{equation}
R = \frac{A}{\tau _{k}} = -c\mbox{Im}\left[ T(\omega) \right]= R_0+R^{eff}_{DG-F},
\label{Eq1} 
\end{equation}
with 
\begin{equation}
A=3\Omega /e^{2}v_{F}\rho_s(\epsilon _F),
\label{A0}
\end{equation} 
where, $\rho _s$ denotes the s-band DOS, $\Omega$ the volume of lanthanum ion host and $v_{F}=\sqrt{2\epsilon _F/m_e}$ the electron velocity at the Fermi's surface.
In equation~(\ref{Eq1}) the effective excitation mean life, $\tau _{k}$, is related to resistivity, $R$, through the imaginary part of the scattering matrix $T(\omega)$ \cite{Blackman}. 
In our extended s,d-band model, the terms in equation (\ref{Eq1}) at the Fermi ($\epsilon _F$) level are expressed as:
\begin{enumerate}[1.]
\item The residual resistivity,
\begin{eqnarray}
R_0(\epsilon _F) & = & Ac\vert V_{sd}\vert ^2\left\{ \frac{\vert F_d(\epsilon _F)\vert   }{\vert   X (\epsilon _F)\vert   } \times \sin [\delta _d(\epsilon _F) - \eta (\epsilon _F)] \right\}, \nonumber \\
& = & \frac{Ac\vert V_{sd}\vert ^2}{\vert X (\epsilon _F)\vert }[\pi \rho _d(\epsilon _F) \cos \eta (\epsilon _F) - F^R_d(\epsilon _F)\sin \eta (\epsilon _F) ], 
\label{DELTAR0} 
\end{eqnarray}
\item The spin resistivity in terms of $R_{DG-F} = \pi A c \rho_s(\epsilon _F )(J^{(s)})^2S_0(S_0+1)$,
\begin{equation}
R^{eff}_{DG-F} = \left\{ 1+\theta (\epsilon _F, \vert V_{sd}\vert ^2)\right\} R_{DG-F}
\Rightarrow J_{eff}^{(s)}=J^{(s)}\left\{ 1+\theta (\epsilon _F,\vert   V_{sd}\vert ^2) \right\} ^{1/2} 
\label{JEFF} 
\end{equation}
\end{enumerate}
with,
\begin{eqnarray}
\theta(\epsilon _F,\vert V_{sd}\vert   ^2) & = & \frac{\vert K(\epsilon _F)\vert ^3}{\vert X (\epsilon _F)\vert ^3} \times \cos \left\{3[\eta _{dd}(\epsilon _F) - \eta (\epsilon _F)]\right\} \label{HExacta} \\
& + & \frac{\vert   K(\epsilon _F)\vert }{\vert X (\epsilon _F)\vert } \frac{\vert   F_s(\epsilon _F)\vert }{F^I_s (\epsilon _F)} \times \sin \left\{3[\eta _{dd}(\epsilon _F) - \eta (\epsilon _F)]\right\} \nonumber \\
& + & 2 \vert V_{sd}\vert ^2 \left( \frac{J^{(d)}}{J^{(s)}} \right) \frac{\vert F_s(\epsilon _F)\vert }{F^I_s (\epsilon _F)} \times \frac{\vert K(\epsilon _F)\vert   \vert F_d(\epsilon _F)\vert ^2}{\vert X (\epsilon _F)\vert ^2} \nonumber \\
& \times & \sin [\eta _{dd} (\epsilon _F) + \delta _s (\epsilon _F) + 2\delta _d (\epsilon _F) -3\eta (\epsilon _F)] \nonumber \\
& + & \vert V_{sd}\vert ^2 \left( \frac{J^{(d)}}{J^{(s)}} \right)^2 \frac{\vert F_d(\epsilon _F)\vert ^3}{\vert X (\epsilon _F)\vert ^3} \left[ \frac{1}{ F^I_s (\epsilon _F) } \right] \nonumber \\ 
& \times & \sin \left\{3[\delta _d (\epsilon _F) - \eta (\epsilon _F)]\right\}. \nonumber 
\end{eqnarray}
In equations~(\ref{DELTAR0}) and (\ref{HExacta}), $F^I_s (\epsilon _F)=\pi \rho _s (\epsilon _F)$, 
$\vert F_{\lambda}\vert $ and $\delta _{\lambda}$ ($\lambda = s,d$), are related to the band structure of the host. In contrast, $\vert K\vert $ and $\vert \eta _{dd}\vert$ belongs to the charge potential introduced by the impurity (see~\ref{Ap1}).
\begin{itemize}
\item The function $\theta $ in equation (\ref{JEFF}) describes the combined effect of charge potential and spin-dependent scattering reflecting the deviations from the De~Gennes-Friedel resistivity~\cite{DF}.
\item The cross-products appearing in the expression of $ \theta $ in equation (\ref{JEFF}), $(J^{(s)})^2$, $(J^{(d)})^2$ and $J^{(s)}\times J^{(d)}$), have significant influence on the magnitude of the effective exchange parameter $J^{(s)}_{eff}$.
\item In our model, the matrix element $\vert V_{sd}\vert ^2$ is treated as a phenomenological parameter.
\item Two primary charge effects introduced by the impurities are incorporated into the present model compared to Ref.~\cite{Atroper}. 
These effects are first, the translational symmetry breaking, which results in a non-local charge potential centered at the d-band \cite{Zeller},
\begin{equation}
V_{dd}^{\star}(\omega)=V_{dd}+(\alpha ^2 -1)(\omega-\epsilon _c ^d)+\vert V_{sd}\vert   ^2 P_{00}^{ss}(\omega),
\label{Vddstar}
\end{equation}
which is calculated from an extended Friedel's condition, by considering the fivefold degeneracy of the d-band as identical sub-bands as,
\begin{equation}
\Delta Z = \frac{10}{\pi }\tan ^{-1} \frac{\pi \vert V_{dd}^{\star}(\epsilon _F)\vert \rho _d(\epsilon _F)}
{1-\vert V_{dd}^{\star}(\epsilon _F)\vert F^R _d(\epsilon _F)},
\label{deltaZ}
\end{equation} 
and second, the excess of charge $\Delta Z$ introduced by the impurity which is is incorporated as,
\begin{equation}
\Delta Z^{\prime } = \Delta Z - \frac{\Omega ^I -\Omega }{\Omega } = \Delta Z - \delta v/\Omega.
\label{VolEf} 
\end{equation}
In equation (\ref{VolEf}), $\Omega ^I$ and $\Omega $ are the volumes of the impurity and hots ions, respectively. Note that, when $\alpha =1$ we recover the results in Ref.~\cite{Atroper}.
\end{itemize}

\section{Computational procedure}
\label{S4}
As discussed in the previous section, various quantities contribute to the resistivity. An important goal of this work is to incorporate values derived from first-principles calculations into the resistivity expression. In this section, we outline the computational procedure used to achieve this. The numerical results will be presented in the following section.
The bulk crystal structure and electrical properties of lanthanum were calculated using the Vienna Ab Initio Simulation Package (VASP) \cite{VASP1, VASP2}. 
We employed the projector augmented wave (PAW) method in these calculations to model the interaction between the potential core and valence electrons with a plane wave cutoff energy set at 520 eV to converge the total
energy of La below $1$ meV per atom. The convergence criterion for electronic minimization is an energy difference smaller than $10 ^{-8}$ eV. 
The calculations were spin-polarized, using the Perdew-Burke-Ernzerhof \cite{VASP3, VASP4} parameterization of the generalized gradient approximation (GGA) for the exchange-correlation potential. 
For the primitive cell, the initial coordinates were sourced from the Materials Project database \cite{MatProj}. During the volume scan procedure, we allowed variations of the positions of the cell's ions, as well as adjustments to its shape and volume, ultimately leading to the determination of the optimal volume for the primitive cell. Careful convergence tests indicated that an $18 \times 18 \times 9$ k-points grid mesh was enough for self-consistent calculations of the lanthanum structure. 
Using this k-point mesh, we drive the relaxation procedure for at least 16 different cell sizes L (the edge length of the supercell), varying by approximately $\pm 0.4\%$ around the size corresponding to minimum energy. A parabolic fit with a third-order Birch-Murnaghan equation determined the minimum energy and its associated volume.
We relax the geometries using conjugate gradient until the force on each atom is less than 5 meV/\AA. Each relaxation procedure allowed fractional electronic occupancies through Gaussian smearing, with a width of $\sigma $ = 0.03 eV. 
The process concluded with a self-consistent cycle, during which Bl\"ochl's methodology for Brillouin zone integration was employed~\cite{BLOCHL}. Brillouin-zone sampling was performed using a Gamma-centered k-point grid, as recommended by the package defaults. 
Defect calculations were performed using a periodic $2 \times 2 \times 1$ supercell containing 15 La atoms and a substitutional impurity. For these calculations, a $9 \times 9 \times 6$ k-point grid was used, maintaining the cell shape and volume fixed to that of pure La while allowing internal ionic relaxations.
The supercell was fully relaxed to ensure that each atom reached its equilibrium position, minimizing the system's total energy. All relaxation procedures were executed using a quasi-Newton method.
We have calculated the density of states with a k-point grid that is twice as dense as before, alongside the band structure, using high-symmetry k-points obtained from the VASPkit utilities~\cite{VASPKIT}, which were verified on the SEEK website~\cite{SEEK}. For the alloys, the energy of solute atoms in a particular site is generally referred to as that of the most stable site of the system. The converged lattice parameters and the calculated band structure of both, the primitive cell and the supercell of lanthanum are now presented.

\begin{figure} 
\begin{center}
\includegraphics[scale=0.3]{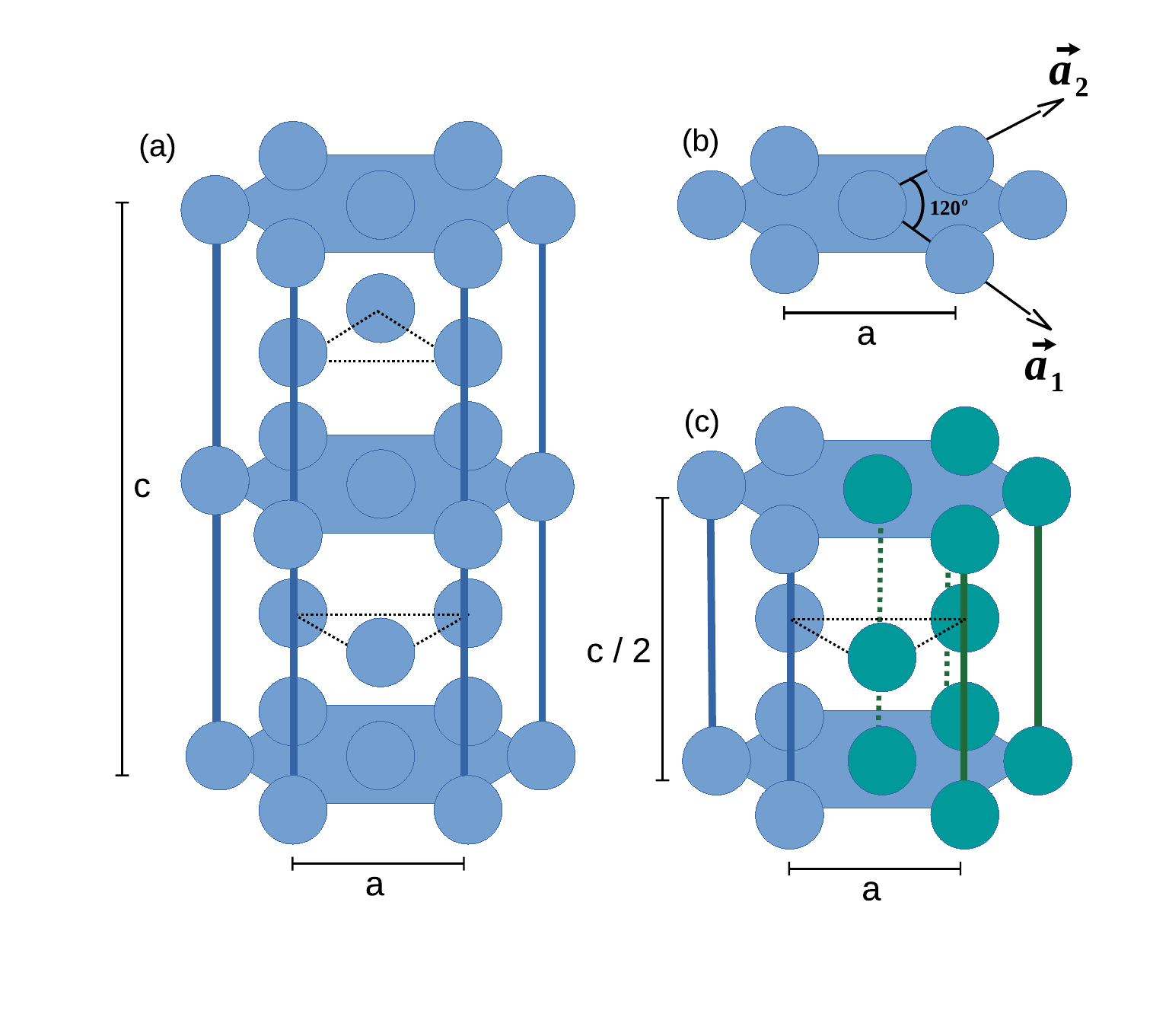} 
\caption{(Color online) (a) Scheme of double hexagonal close-packed structure of La. In (b) the primitive vectors and (c) the primitive cell (in green).} 
\label{FIG1} 
\end{center}
\end{figure}

\begin{figure} 
\begin{center}
\includegraphics[scale=0.62]{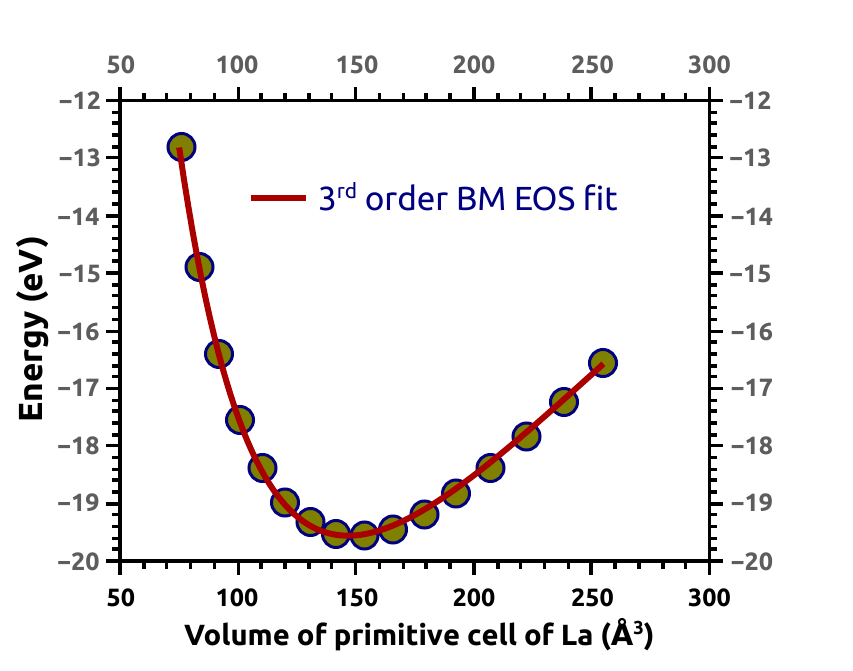}
\caption{(Color online) Energy (in eV) vs Volume of the primitive cell (in \AA$^3$) from VASP calculations (green circles). The Birch-Murnaghan third-order equation of state fit (red line). } 
\label{FIG2} 
\end{center}
\end{figure}

The primitive cell depicted in figure~\ref{FIG1}, containing four atoms of La, is obtained considering $n_x=n_y=n_z=1$ in equation (\ref{VP}). We use the generalized gradient approximation GGA as parameterized by the PBE scheme~\cite{VASP3, VASP4}.
\begin{equation}
dhcp-La=\left\{
\begin{array}{ll}
\vec{a}_1 = \frac{a}{2}n_x\hat x - \frac{\sqrt{3}a}{2}n_y\hat y \\
\vec{a}_2 = \frac{a}{2}n_x\hat x + \frac{\sqrt{3}a}{2}n_y\hat y\\
\vec{a}_3 = c n_z \hat z
\end{array}
\right. 
\label{VP}
\end{equation} 
The third-order Birch-Murnaghan equation of states (EOSs) was used to fit the calculated energy-volume (E-V) data for the primitive cell of lanthanum. The results are depicted in figure~\ref{FIG2}.
After the scan volume calculations, we relaxed the crystal structure with the lattice parameters obtained from the E-V fit. Following this procedure, we get the optimized lattice parameters $a=3.774$ \AA, $c=12.084$ \AA, and the ratio $c/a=3.202$ (or $c/2a=1.601$) for lanthanum. Also, the La ion results in $\Omega =37.15$ \AA$^3$ and formation energy $E_0=-4.90$ eV per atom. 
Table~\ref{TAB1} summarizes the optimized lattice parameters of lanthanum, including the ratio $c/a$ and the lattice constant $a$. Additionally, the corresponding ionic volume $\Omega $ and the relative total energy per atom $E_0$ are also provided. Results for the supercell of La containing 16 and 36 atoms are also displayed in the same table.
\begin{table}
\caption{\label{TAB1} Optimized lattice parameters of lanthanum ion host and the rare-earth impurities, $c/a$ and $a$; the corresponding volume $\Omega $ and $\Omega ^I$, respectively and the total energies $E_0$ per atom.} 
\centering
\begin{tabular}{lcccc}
\hline
Host  & $c/a$ & $a$(\AA) & $E_0$(eV) & $\Omega$(\AA)$^3$ \\
\hline
La$_4$              & 3.202 & 3.774 & -4.89 & 37.15 \\ 
Ref.\cite{SCHOLL11} & 3.206 & 3.754 &       & 36.71 \\
Ref.\cite{LAEXP90}  & 3.225 & 3.774 &       &       \\
\hline
La$_{16}$           & 3.206 & 3.777 & -4.89 & 37.18 \\
La$_{36}$           & 3.208 & 3.776 & -4.89 & 37.18 \\
\hline
\hline
Imp.  & $c/a$ & $a$(\AA) & $E_0$(eV) & $\Omega ^I$(\AA)$^3$ \\
\hline
Gd$_3$ & 1.567 & 3.646 & -4.59 & 32.60 \\
Tb$_3$ & 1.555 & 3.638 & -4.56 & 32.50 \\
Dy$_3$ & 1.550 & 3.622 & -4.84 & 31.99 \\
Lu$_3$ & 1.558 & 3.517 & -4.46 & 29.43 \\
\hline
\end{tabular}   
\end{table}
As shown in Table~\ref{TAB1}, our results are in well agreement with the calculated values $c/a= 3.206$, $a=3.754$, and $\Omega=36.71$ reported by Sch\" ollhammer \textit{et al.} in Ref.~\cite{SCHOLL11} and with the experimental data of $c/a=3.225$ for $a=3.774$ in Ref.~\cite{LAEXP90}. Also, the present results are in very good agreement with the calculated values of Lang in Ref.~\cite{LANG17,LANG21}. The results for the impurities are in excellent agreement with results in the Materials Project site \cite{MatProj}. 

It is important to note that, our model considers the more realistic situation where the Fermi energy $\epsilon _F$, the energy centers $\epsilon ^{s} _c$ and $\epsilon ^{d} _d$ of the s- and d-bands, as well as, their bandwidths, $\Delta _{s}$ and $\Delta _{d}$, are not null. Table~\ref{TAB2} summarizes these magnitudes from present DFT calculations using the Vaspkit utilities~\cite{VASPKIT} in the energy range of $\left[-5,15\right]$ eV.

\begin{figure} 
\begin{center}
\includegraphics[scale=0.62]{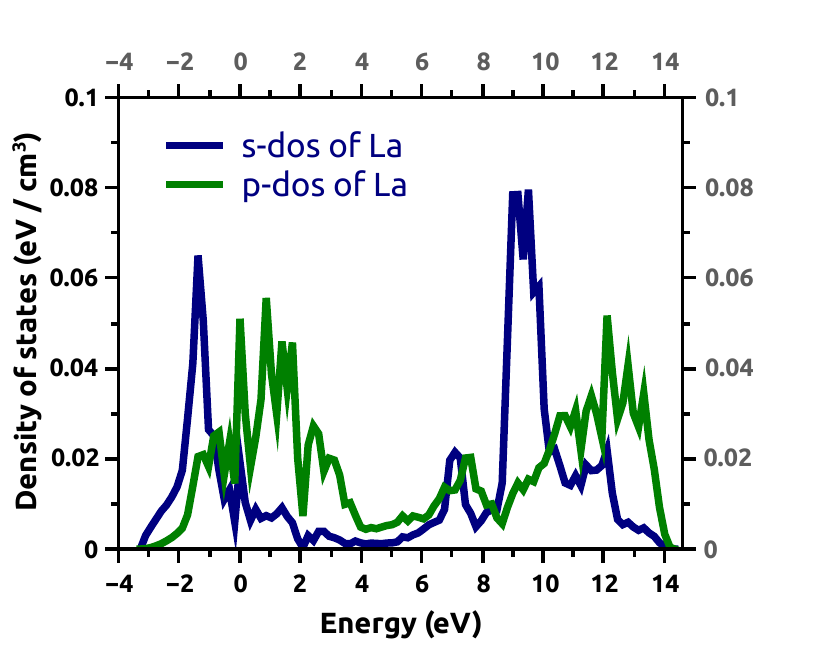} 
\includegraphics[scale=0.62]{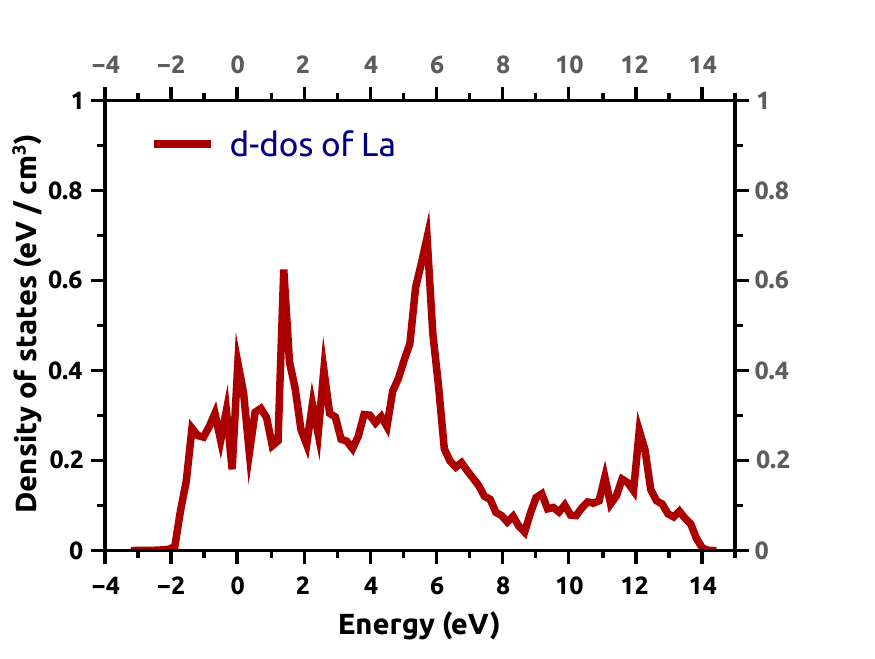} 
\includegraphics[scale=0.62]{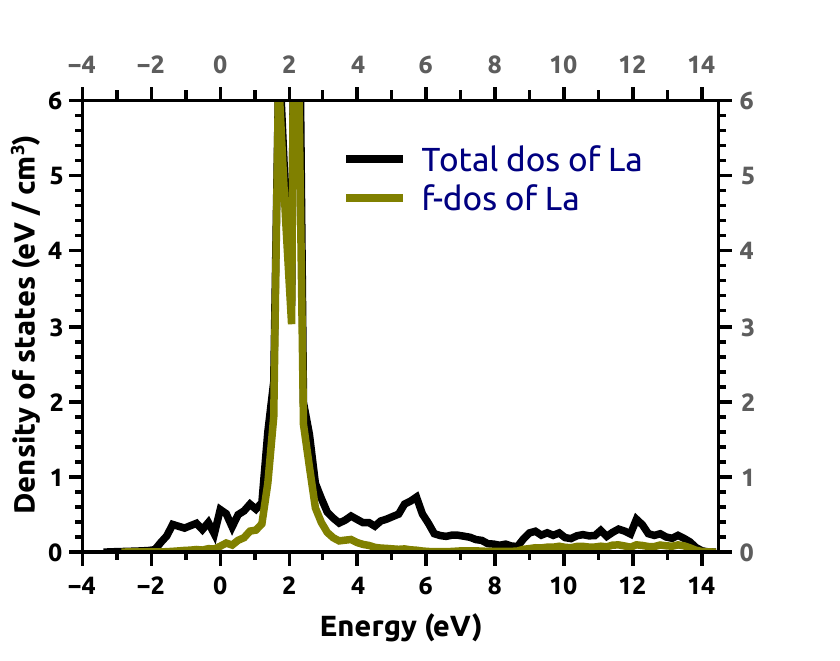} 
\caption{Density of states of lanthanum (in eV/cm$^2$) vs energy (in eV)): Top panel, DOS for s and p bands in blue and green, respectively. Middle panel, DOS of the d band. Bottom panel, the total (in black) and f band DOS (in olive).} 
\label{FIG3} 
\end{center}
\end{figure}

\begin{figure} 
\begin{center}
\includegraphics[scale=0.36]{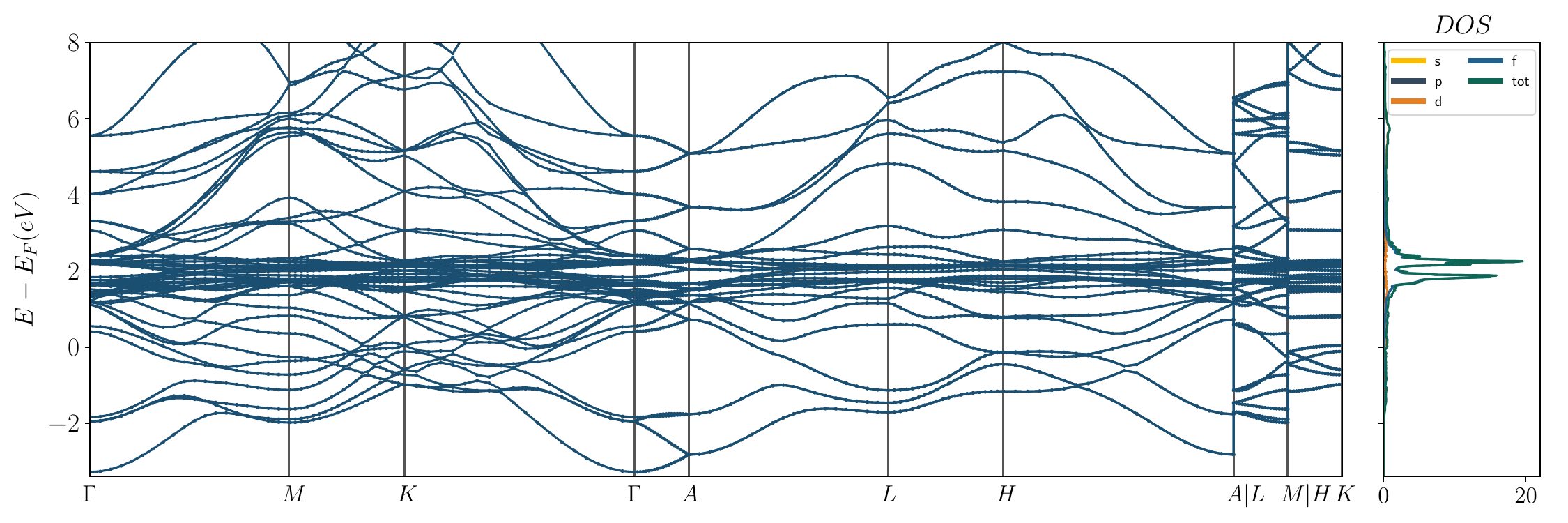}
\caption{Band structure and density of states of La.} 
\label{FIG4} 
\end{center}
\end{figure}

\begin{table}
\caption{\label{TAB2} Parameters involved in the calculation of $A$ in equation (\ref{A0}). The Fermi energy $E_f$ (in eV), the centers of the s- and d-bands $\epsilon ^{s} _c$ and $\epsilon ^{d} _d$, and the bandwidths $\Delta _s$ and $\Delta _d$ for lanthanum. Also the Fermi electron velocity $v_F$ (in $\times 10^{6}$ m/s).} 
\centering
\begin{tabular}{llcccccc}
\hline 
 Host & $\epsilon _F$ & $\epsilon ^s_c$ & $\epsilon ^d_c$ & $\Delta _s$ & $\Delta _d$ & $v_F$ ($\times 10^{6}$ m/s) & Refs.  \\
\hline
 La$_4$       & 7.83 & 6.35 & 4.62 & 17.29 & 17.12 & 1.66 & DFT \\
\hline
\end{tabular}  
\end{table}
Since lanthanum is a non-magnetic material, in Figure \ref{FIG3} we only show the density of states for the spin up polarization. The DOS for the s- and p-bands are shown together in the top panel in blue and green, respectively. While the middle panel depicts the DOS for the d-band (in red). At the bottom of the figure, the total (in black) and f-band (in olive) DOS are displayed. 
Since our calculations were spin polarized, in Fig.\ref{FIG4} we summarizes the bands structure and DOS for both spin polarization for lanthanum host. We can then observe the symmetric behavior characteristic of non-magnetic materials.
Our results in Figs.\ref{FIG3} and \ref{FIG4} are in well agreement with those in Ref.\cite{LIN21} for GGA calculations.

In Addition, we have calculated the unfolding band structure of pure lanthanum containing 16 atoms and the density of states of the four La alloys, specifically La${15}$Gd, La${15}$Dy, La${15}$Tb, and La${15}$Lu. The calculations were performed for a supercell of 16 atoms, constructed from the coordinates of the primitive cell of La$_4$, using Vaspkit~\cite{VASPKIT} with parameters $n_x = n_y = 2$ and $n_z = 1$. In these alloys, the impurities are incorporated at substitutional sites. These figures are summarized as Supplementary data.

In the next section we calculate the resistivity for the here described DOS band model as well as for a parabolic band case.

\section[Numerical results]{Numerical results}
\label{S5}

In this section we calculate numerically the resistivity for La diluted alloys comparing two DOS band models, the more realistic one described in the previous section and the parabolic band model which is a simplified approximation vastly used in the literature. 

The parabolic density of states is defined as follows:
\begin{equation}
\rho_{\lambda }(\epsilon ) = \alpha _{\lambda} (\epsilon ^2- \Delta ^2_{\lambda}) \,\,\, \mathrm{(\lambda = s \, or \,  d)}, 
\label{Eq22}
\end{equation}
where $\alpha _{\lambda}$ are normalization constants, taken as $0.75 \Delta ^{-1} _{\lambda}$ to ensure one electron per atom, $\Delta _{\lambda}$ represents the corresponding energy width as in Ref.\cite{Troper78}. On the other hand, our DFT calculations lead to $\Delta _{d}/\Delta _{s}= 0.88$ for dhcp La. The density of states $\rho_{\lambda }(\epsilon )$ for parabolic as well as for DFT calculations corresponds to a band fillings of five electrons.

The parameters involved in the expressions (\ref{DELTAR0}) -(\ref{HExacta}) include the constant $A=3\Omega /e^{2}v_{F}\rho_s(\epsilon _F)$. These set of values are obtained from our DFT calculations and they are summarized in Table\ref{TAB2} as noted in the previous section.  

Concerning the Spin $S_0$ impurity involved in equation (\ref{JEFF}), we assume that the rare-earth impurity behaves as a ``spherically symmetric'' entity; therefore, we disregard effects such as skew scattering (related to spin-orbit coupling) and crystal field splitting. 
Consequently, our calculations most apply to the Gd impurity in transition metal-like hosts, where the total spin is $S_0(Gd)=7/2$.
However, we can also consider impurities such as Tb and Dy, which have non-zero orbital angular momentum ($L \neq 0$). In these cases, we replace $S_0$ with the effective spin $(g_j-1)J_0= S_0^{eff}$, where $(g_j-1)$ is the De~Gennes factor that represents the projection of the spin $S_0$ onto the total angular momentum $J_0$ of Tb and Dy impurities. The corresponding electronic configuration and $S_0$ values are shown in Table \ref{TAB3}.
\begin{table} 
\caption{\label{TAB3} Electronic configuration and $S_0$ values for Gd, Tb, Dy, and Lu, as adapted from Ref.~\cite{Kobler10}.} 
\centering
\begin{tabular}{llcc}
\hline 
 Imp. & Elec. Config. & $S_0^{eff}$  \\
\hline 
Gd    & $[Xe]4f^75d^16s^2$    & 7/2  \\
Tb    & $[Xe]6s^24f^9$        & 6/2    \\
Dy    & $[Xe]4f^{10}6s^2$     & 5/2  \\
Lu    & $[Xe]4f^{14}5d^16s^2$ &  0   \\
\hline
\end{tabular}  
\end{table}
The exchange parameters are crucial magnitudes when studying the influence of band structure on the spin resistivity. The magnitude and sign of $J_{eff}$ in the expression of the effective De-Gennes - Friedel resistivity, $R_{DG-F}^{eff}$, in equation (\ref{JEFF}) depend on the extent of mixing between the 4f electrons and the conduction electrons, as well as on the nature of the conduction band in the host material. 
For rare-earth impurities, the exchange interactions are typically described using an indirect exchange mechanism~\cite{CAMP72}.
Due to the localized nature of the 4f electronic shell, the interaction between the 4f spin and an itinerant electron spin occurs primarily through local exchange interactions on the rare-earth atom. 
These interactions are described by the intra-atomic exchange integrals $J_{4f-5d}$, $J_{4f-6s}$ and $J_{4f-6p}$, with the $J4f-5d$ integral being the dominant term \cite{CAMP72}.
Then, we use the exchange integrals $J_{4f-6s}$, $J_{4f-6p}$ and $J_{4f-5d}$ from relativistic atomic calculations obtained in Ref.~\cite{LI91} and shown in Table \ref{TAB4}. 
From this table, we derive the necessary ratios $J^{(d)}/J^{(s)}=5$ and $J^{(s)}/J^{(d)}=2$ that appear in equation (\ref{HExacta}).
\begin{table}
\caption{\label{TAB4} The impurity exchange parameters (in eV).} 
\centering
\begin{tabular}{lccc}
\hline
Impurity & $J_{4f-5d}$ & $J_{4f-6s}$ & $J_{4f-6p}$  \\
\hline
Gd$^{3+}$ & 0.1891 & 0.0377 & 0.0151 \\
Tb$^{3+}$ & 0.1837 & 0.0375 & 0.0148 \\
Dy$^{3+}$ & 0.1797 & 0.0374 & 0.0145 \\
\hline
\end{tabular}  
\end{table}
Experimental results of the resistivity for both fcc and hcp lanthanum (La), containing small amounts of rare earth (RE) impurities such as gadolinium (Gd), are sourced from Ref. \cite{SUGA66}. Additionally, the effective s-f exchange integral $J_{eff}^{(s)}$ has been calculated based on the resistivity data. In Ref. \cite{SUGA66}, the authors compared $J_{eff}^{(s)}$ with values derived from changes in the superconducting transition temperature $\Delta T_c$, as described by Abrikosov and Gor'kov \cite{ABRIK61}. There, $\Delta T_c$ is defined as $T_c(La-1\%Re) - T_c(La)$, where $T_c$ is the critical temperature for the fcc La-RE alloys containing 1 atomic percent of RE and the pure $T_c$ of fcc La. The effective exchange integrals obtained in Ref. \cite{SUGA66} are $J_{eff}(La-1\%Gd)=0.047$ eV, $J_{eff}(La-1\%Tb)=0.066$\,eV and $J_{eff}(La-0.93\%Dy)=0.087$\,eV.
Concerning the impurity valence states of Gd, represented as $4f^{7} 5d^1 6s^2$, it can be expressed in the form $4f^{n-\delta} 5d^{m+\delta} 6s^2$, where $\delta$ denotes the amount of charge transferred from the 4f resonance to the d conduction band. Given that the Gd valence state falls within the range [3.2-3.4], we self-consistently determined $\delta = 0.125$. 
Additionally, we have included the volume difference introduced by the impurity, $\delta v / \Omega $ (as defined in equation (\ref{VolEf})) as well as the energy difference $\Delta E=E(La-I)-E(La)$, where I represents the impurity species (Gd, Tb, Dy, and Lu). Here, $E(La-I)$ is the energy of the supercell of La containing one impurity atom, while $E(La)$ is the energy of pure La host. In present calculations the strongly localized semi-core $f$ electrons in Gd$_3$, Tb$_3$, and Dy$_3$ are treated as core states, even though they are higher in energy than other valence states.
Concerning the non-local charge potential $V_{dd}^{\star}(\omega)$ in equation (\ref{Vddstar}) which enters in both terms of the resistivity, it is obtained from the extended Friedel's condition in equation (\ref{deltaZ}) as explained in details in \ref{Ap2}. 
Table \ref{TAB5} shows our DFT calculations of $\Delta Z=Z_I-Z_h$, needed for such Friedel condition. Note that $\Delta Z$ was determined as the difference between the host and impurity occupation numbers $<n_s>$ and $<n_d>$ (with $<n_d> = 0.1768$ for La). \\
\begin{table}
\caption{ \label{TAB5} The impurity-host charge difference $\Delta Z=Z_I-Z_h$, the occupation difference $\Delta <n>$, the volume difference $\delta v / \Omega $ between the impurity and the La host, and the energy difference $\Delta E$ between the pure host and the alloy.} 
\centering
\begin{tabular}{llccccc}
\hline 
Host  & & & Gd$^{3+}$ &  Tb$^{3+}$ & Dy$^{3+}$ & Lu$^{3+}$ \\
\hline
 T & Elec. Config. & $v_{ion}$ (\AA$^3$) & $\delta v / \Omega $ & $\delta v / \Omega $  & $\delta v / \Omega $ & $\delta v / \Omega $ \\
\hline 
 La  & $ [Xe]5d^16s^2 $  &  37.15 & -0.1282 & -0.1395 &-0.1419 &-0.2106 \\
\hline
  & & $\Delta E$ & -0.28 & -0.30 & -0.34 & -0.42  \\
  & & $\Delta Z$ & -0.89 & -0.79 & -0.69 & -0.77  \\
\hline
\end{tabular}  
\end{table}

\subsection{Critical potential and bound states}
\label{ss1}

The existence of bound states is determined by a critical value, $V_{dd}^c$, calculated as $V^{c}_{dd}= \left[F^R_d(E_{top})\right]^{-1}$ as in eq.(\ref{F11}) of \ref{Ap2}, where $E_{top}$ represents the energy at the top of the d-band.
The limiting value of $V^c_{dd}$ is obtained by adjusting the position of the Fermi level within the band structure to explore the vicinity of the top of d-band. This value depends on the particular choice of the model band shape.
To discuss the role of the band structure in resonant scattering (virtual bound states), we adopt a realistic band structure and a ``parabolic'' DOS model to compare our resistivity results. This approach highlights the effects of changes in the d-band shape of transition metals.

Prior to calculating the resistivity, it is essential to determine the critical potential, $V_{dd}^c$, beyond which bound states begin to emerge. Once $V_{dd}^c$ is obtained, our aim is to examine how resistivity $R$ depends on the densities of states, particularly as a function of the Fermi energy $\epsilon _{F}$.

\begin{figure}[!htbp]
\begin{center}
\includegraphics[scale=0.45]{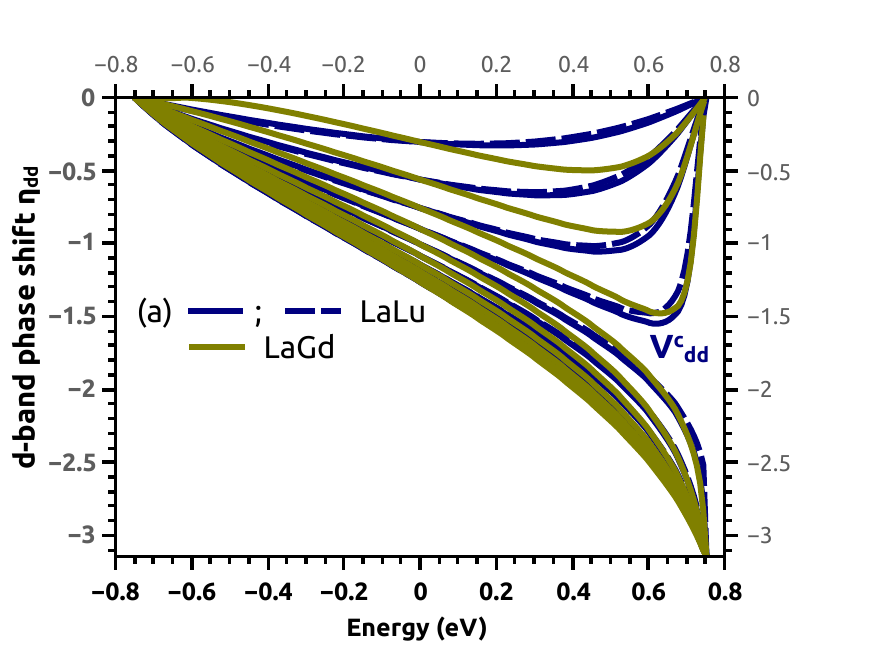} \includegraphics[scale=0.45]{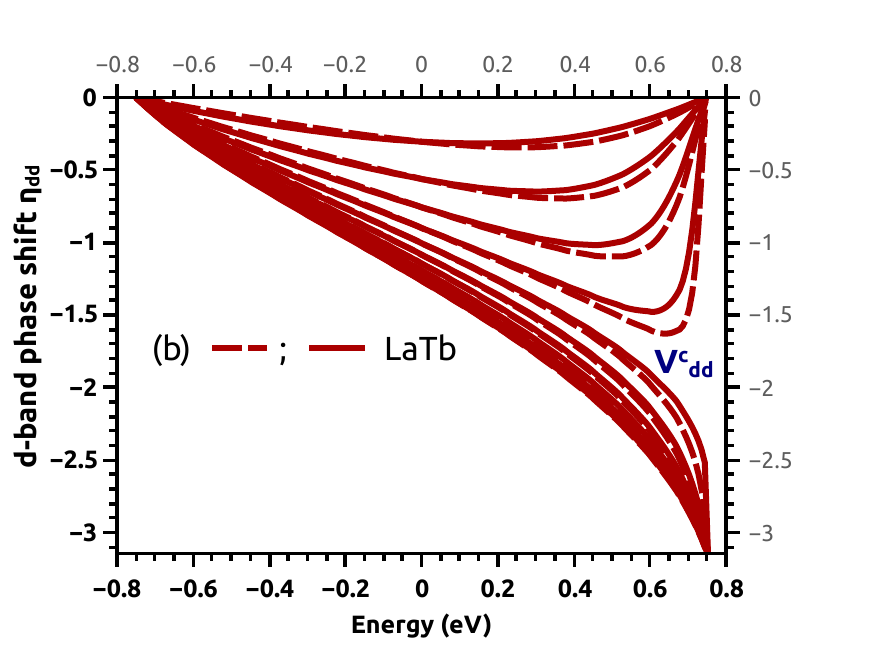} \\
\includegraphics[scale=0.45]{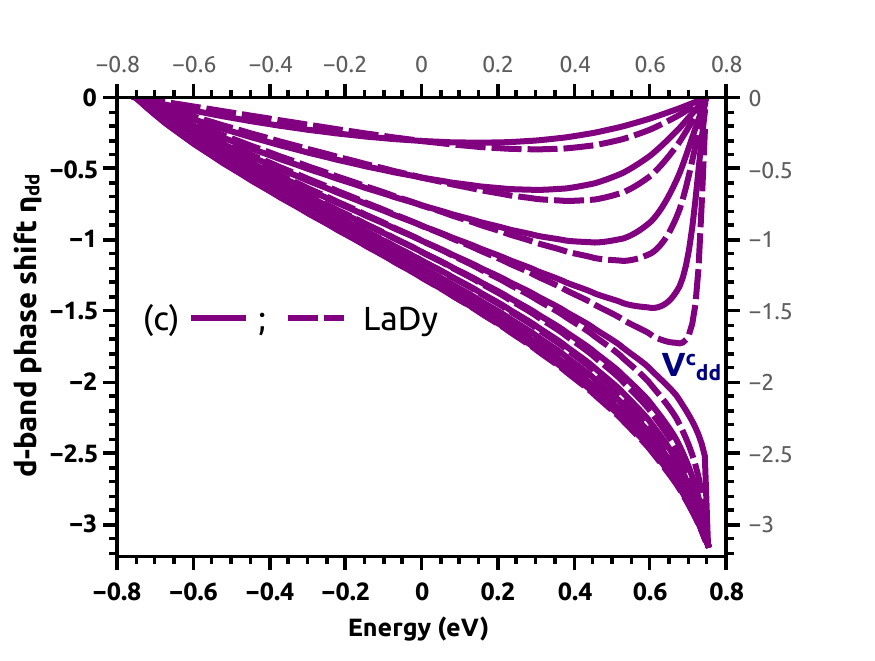}
\caption{Phase shift $\eta_{dd}$ as a function of the energy from the ``parabolic'' band model for lanthanum alloys. Each curve corresponds to a different value of $V_{dd}$ charge potential. In solid lines the case $\alpha \neq 1$ and $\delta v/\Omega \neq 0 $, while dashed lines stand for $\alpha = 1$ and $\delta v/\Omega = 0 $, where period and volume effects were not considered. In (a) our results for La-Gd and La-Lu; (b) La-Tb and (c) La-Dy alloys. Note that the value of $V_{dd}$ that corresponds to the critical potential, $V_{dd}^c$, is highlighted in all panels in red.} 
\label{FIG6}
\end{center}
\end{figure}

\begin{figure} 
\begin{center}
\includegraphics[scale=0.45]{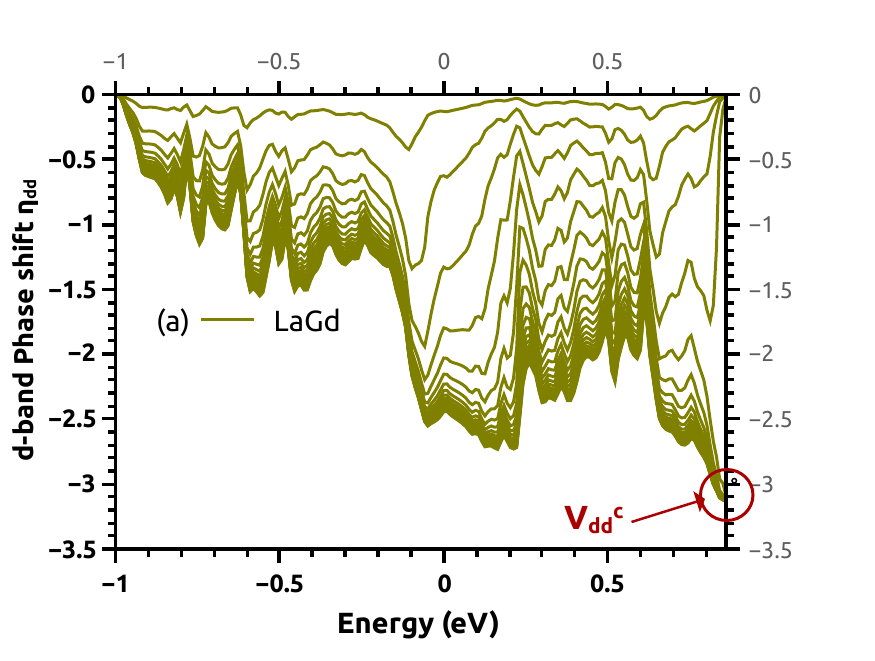} \includegraphics[scale=0.45]{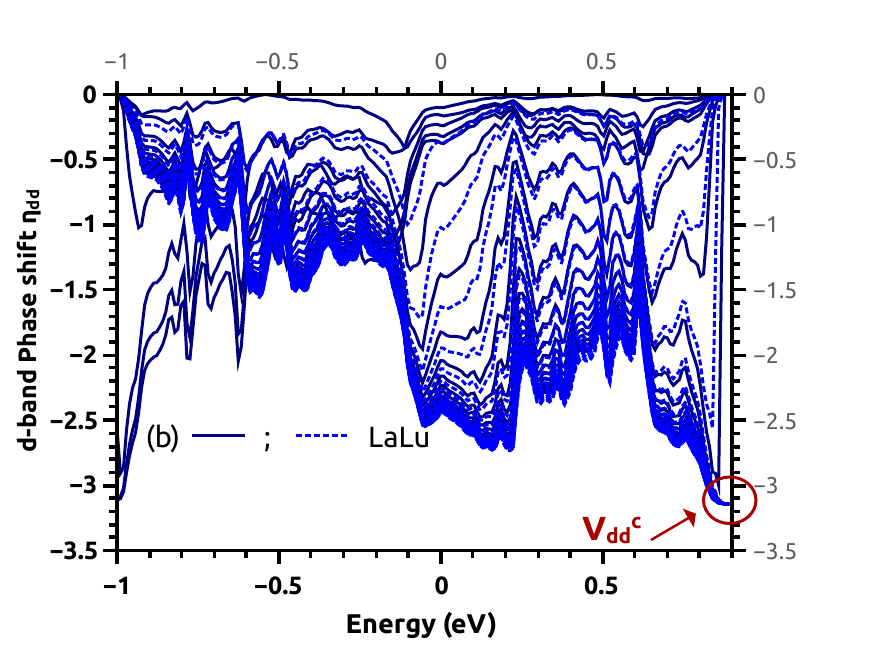}  
\includegraphics[scale=0.45]{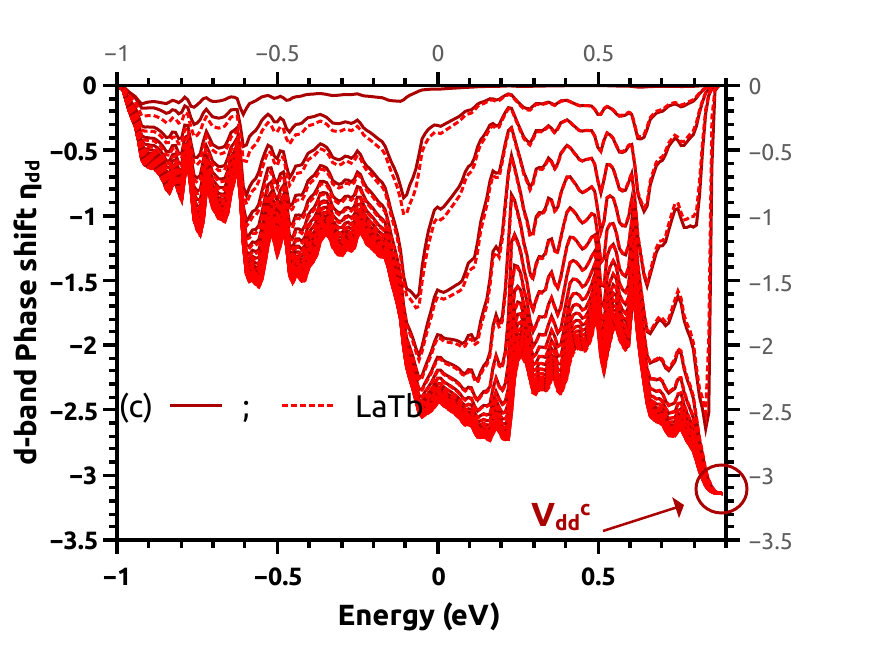} \includegraphics[scale=0.45]{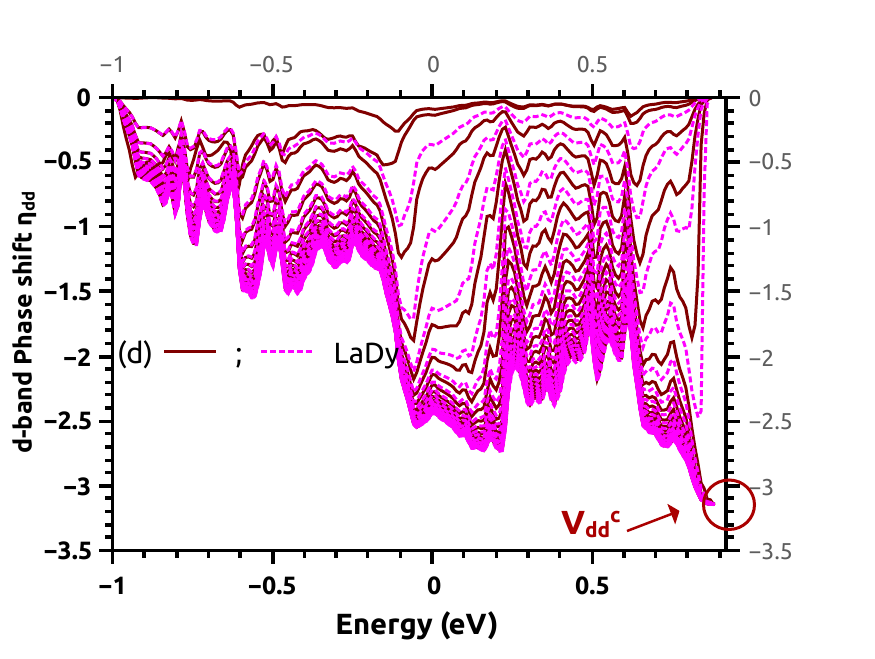} 
\caption{Phase shift $\eta_{dd}$ as a function of the energy from the DFT band model for La-Gd. Each curve corresponds to a different value of $V_{dd}$ charge potential. In solid lines the case $\alpha \neq 1$ and $\delta v/\Omega \neq 0 $, while dashed lines stand for $\alpha = 1$ and $\delta v/\Omega = 0 $, where period and volume effects were not considered. In (a) our results for La-Gd; (b) La-Lu; (b) La-Tb and (c) La-Dy alloys. Note that the value of $V_{dd}$ that corresponds to the critical potential, $V_{dd}^c$, is highlighted in all panels in red.} 
\label{FIG7}
\end{center}
\end{figure}

The calculated $V_{dd}^c$ as a function of  $\epsilon _{F}$ is highlighted for both the ``parabolic'' and DFT models in figures \ref{FIG6} and \ref{FIG7}, respectively. In these figures we show the phase shift, $\eta _{dd} (\epsilon _F)$, as a function of the Fermi energy for different values of the potential $V_{dd}$ ranging from 0.05 eV to 3.0 eV,
\begin{equation} 
 \eta _{dd}(\epsilon) = \cos ^{-1} \frac{1-\vert V_{dd}^{\star}(\epsilon)\vert F_d^{R} (\epsilon) }{[1-V_{dd}^{\star}(\epsilon)F_d ^R(\epsilon)]^2+[V_{dd}^{\star}(\epsilon)F_d ^I (\epsilon)]^2} \, . 
\label{F11a} 
\end{equation}
From which we extract the critical potential value as $V^{c}_{dd}= \left[F^R_d(E_{top})\right]^{-1}$ in eq.(\ref{F11a}) implying in a change of the behavior of $\eta _{dd}$  as a function of the energy.

In figures \ref{FIG6} and \ref{FIG7} we consider two cases. First, in solid lines where $\alpha \neq 1 $ and $\delta v/ \Omega \neq 0$, and secondly dotted lines standing for $\alpha = 1 $ and $\delta v/ \Omega = 0$. The first case represents an extended model compared to that in Ref.~\cite{Atroper}, where the authors have not considered neither period effect (i.e., $\alpha = 1$) nor volume difference between impurity and host (i.e., $\delta v/ \Omega = 0$). 

For the ``parabolic'' model, the critical value for which $\eta _{dd}$ changes its type of behavior occurs consistently at $V_{dd}^{c}=0.683$ eV, for all alloys as shown in Fig.\ref{FIG6}.
 
It is important to recall that, our DFT band model considers the more realistic situation where the center of the s-band $\epsilon _c ^{s}$, the d-band $\epsilon _c^{d}$, and the energy of the Fermi level $\epsilon _F$ are not null. As already mentioned, these parameters are calculated for the DFT band model using VASPKit utilities \cite{VASPKIT} and they are summarized in Table \ref{TAB2}. 

In the case of DFT band model, although the behavior of $\eta _{dd}$ as a function of the energy differs significantly with respect to the ``parabolic'' case, the resulting critical potential value was $V_{dd}^c=0.833$ eV for all La alloys, which is higher than the critical potential obtained for the parabolic band model ($V_{dd}=0.683$ eV). 

We have corroborated that the critical potential occurs at same $V_{dd}$ value for all alloys indistinctly for both the ``parabolic'' and DFT band models. 


\subsection{Resistivity of rare-earth in lanthanum}
\label{ss2}

Up to now, we have obtained all the magnitudes involved for the calculation of the resistivity $R$ for both the parabolic and the DFT band models (see equations (\ref{DELTAR0}) to (\ref{HExacta})). That is, we have obtained the s- and d- band density of states, $\rho_s$ and $\rho_d$, the related bandwidths, $\Delta _s $ and $\Delta _d$, and the energy of the Fermi level, $\epsilon = \epsilon _F$ where the resistivity is evaluated. Also, the $V_{dd}$ scattering matrix element has been derived using an extended Friedel's sum rule (see \ref{Ap2}) and the exchange parameters $J^{(s)}$ and $J^{(d)}$ in Table\ref{TAB4}.

Once this has been performed, we can finally present our results for the temperature-independent resistivity, $R$.

\begin{figure} 
\begin{center}
\includegraphics[scale=0.42]{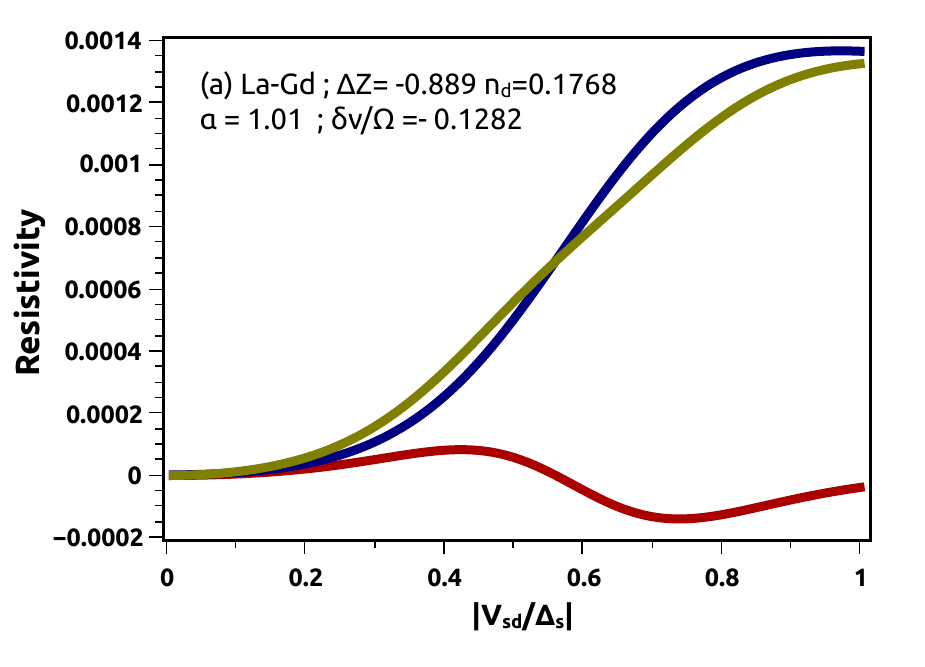} \includegraphics[scale=0.42]{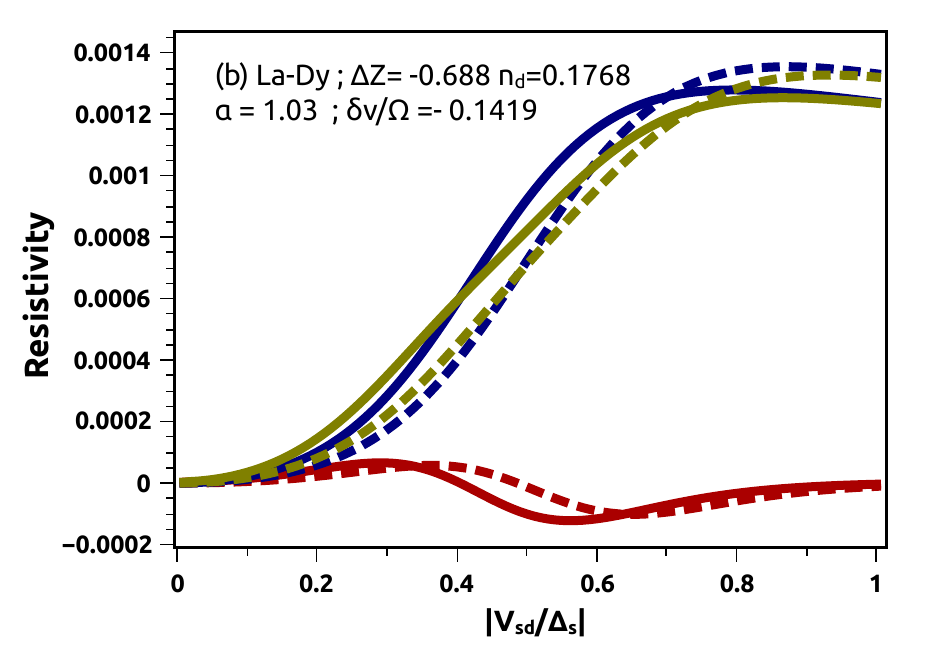} \\
\includegraphics[scale=0.42]{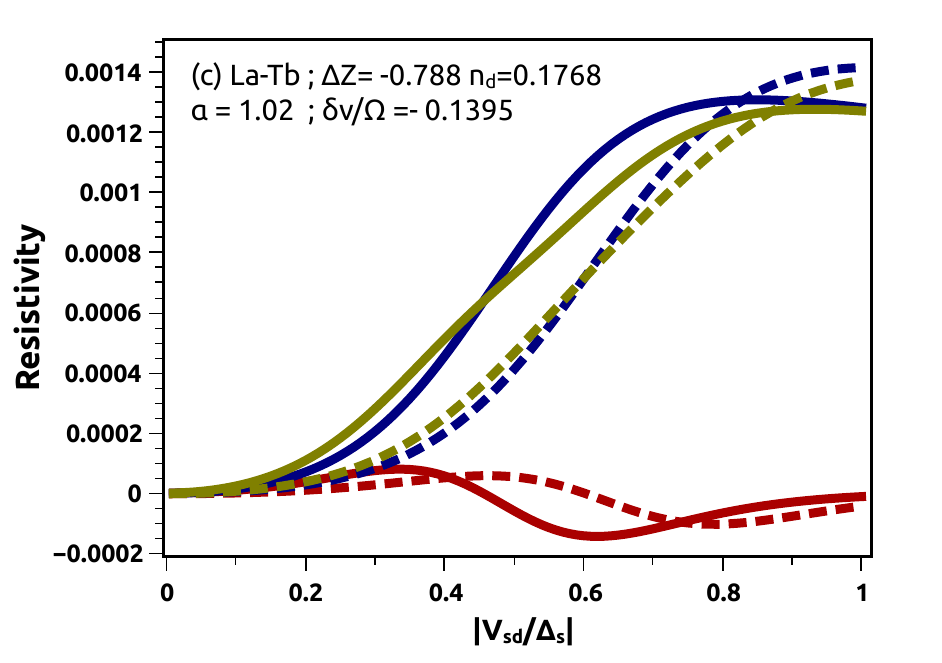} \includegraphics[scale=0.42]{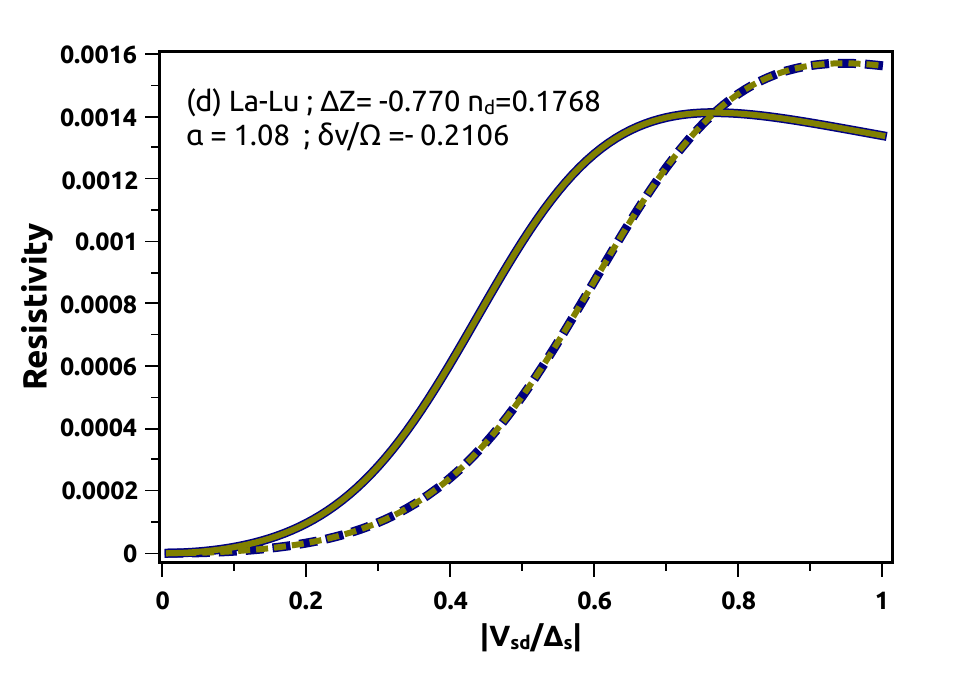}
\caption{Resistivity of rare-earth impurities in La in terms of $\vert V_{sd}/\Delta _s\vert $ for the ``parabolic'' band model. (a) La-Gd, (b) La-Dy, (c) La-Tb and (d) La-Lu alloys.  With, the total resistivity in green, the residual resistivity in blue and the spin dependent resistivity in red lines, respectively.} 
\label{FIG12} 
\end{center}
\end{figure}

In figures \ref{FIG12} and \ref{FIG14}, solid lines represent cases where $\alpha \neq 1$ and $\delta v/\Omega \neq 0 $, while dotted lines correspond to $\alpha = 1$ and $\delta v/\Omega = 0 $, as described in equations (\ref{Veff2}) and (\ref{VolEf}), as described earlier. 

In figure \ref{FIG12} we present our results for the resistivity calculated using the ``parabolic'' band model as described in equation (\ref{Eq22}), respectively for Gd, Tb, Dy and Lu diluted in La.  For this alloys, the figure \ref{FIG12} shows in each panel the total resistivity in green, the residual resistivity in blue and the spin dependent resistivity in red lines, respectively.

Note that in Fig.\ref{FIG12}(a) for the La-Gd alloy, we were unable to achieve convergence for Friedel’s sum rule at $\alpha =1$ and $\delta v/\Omega =0$ across any of the parameter ranges studied.

For the parabolic model, the spin resistivity, $R_{DG-F}$ (in red), can take positive and negative values. This behavior significantly impacts the total resistivity above $\vert V_{sd}/\Delta _s\vert > 0.1$ on (a) La-Gd, (b) La-Dy, and (c) La-Tb alloys, as shown also in figure \ref{FIG12}. In same figure, in the case of (d) La-Lu alloys, although there is no spin contribution to the total resistivity, a considerable difference in $R=R_0$ is observed with the period and volume effects for $\vert V_{sd}/\Delta _s\vert > 0.1$. 

\begin{figure} 
\begin{center}
\includegraphics[scale=0.46]{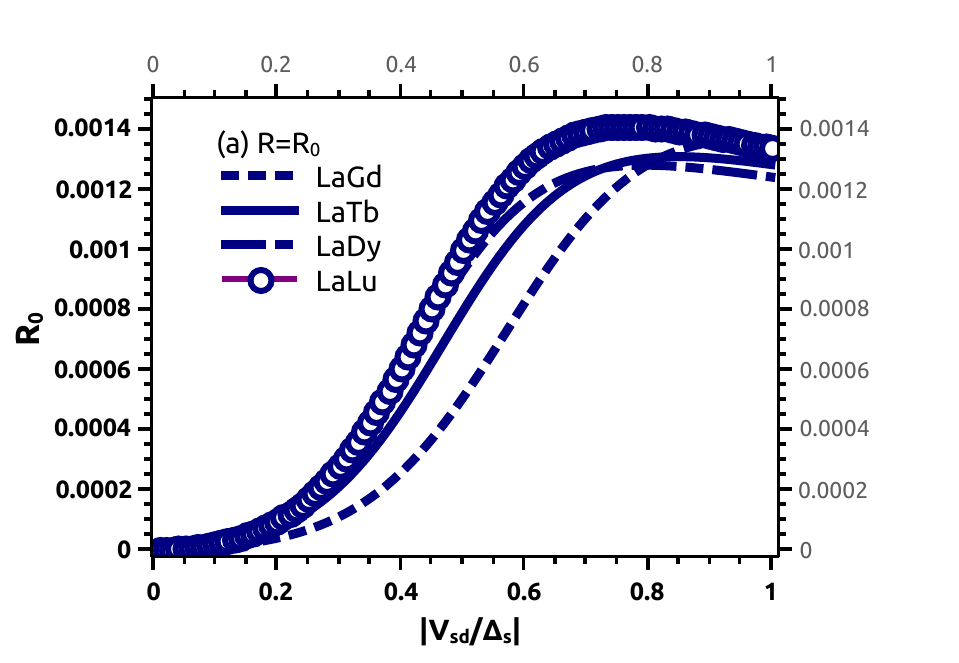} \includegraphics[scale=0.46]{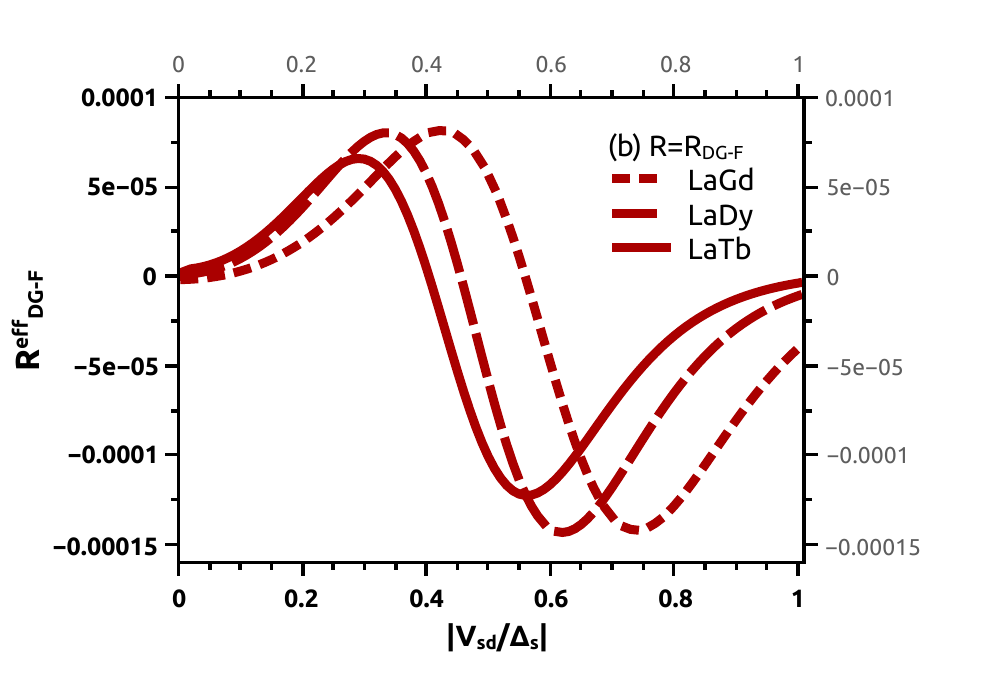} \\
\includegraphics[scale=0.46]{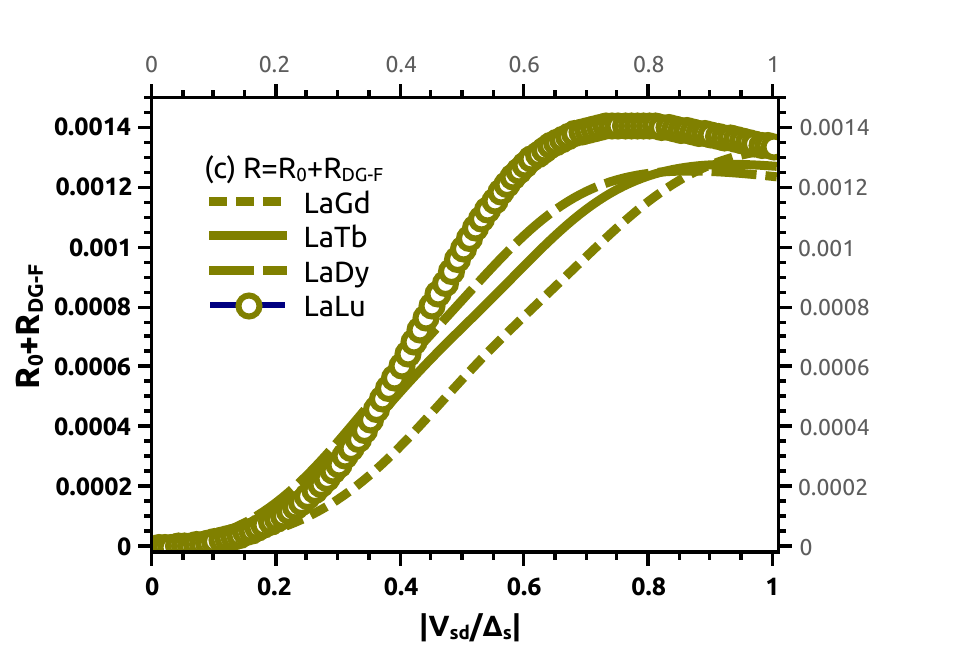} 
\caption{Comparison of the (a) residual resistivity $R_0$ (in blue), (b) the spin resistivity $R_{DG-F}^{eff}$ (in red), and (c) the total resistivity $R=R_0+R_{DG-F}^{eff}$ (in green), in terms of $\vert V_{sd}/\Delta _s\vert $ using the ``parabolic'' band model. The solid, dotted, dashed and circles lines correspond to La-Tb, La-Gd, La-Dy and La-Lu respectively.} 
\label{FIG13} 
\end{center}
\end{figure}
\begin{figure} 
\begin{center}
\includegraphics[scale=0.43]{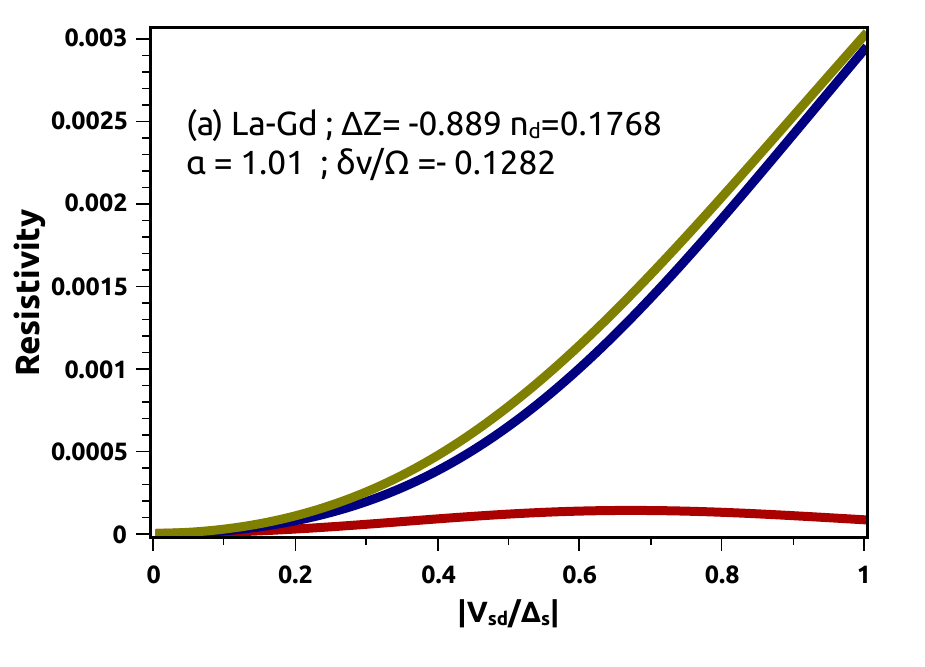} \includegraphics[scale=0.43]{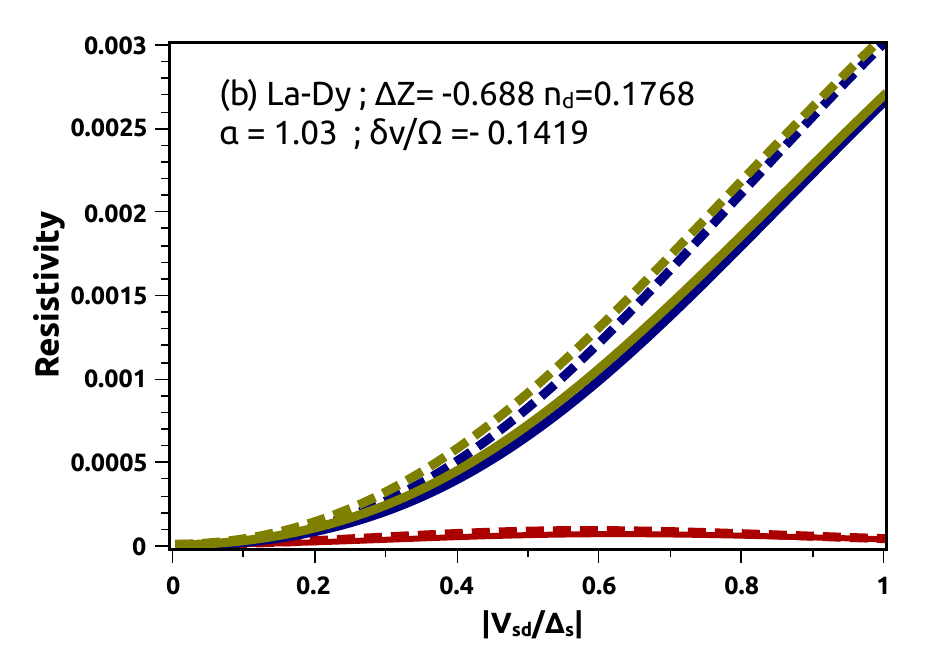} 
\includegraphics[scale=0.43]{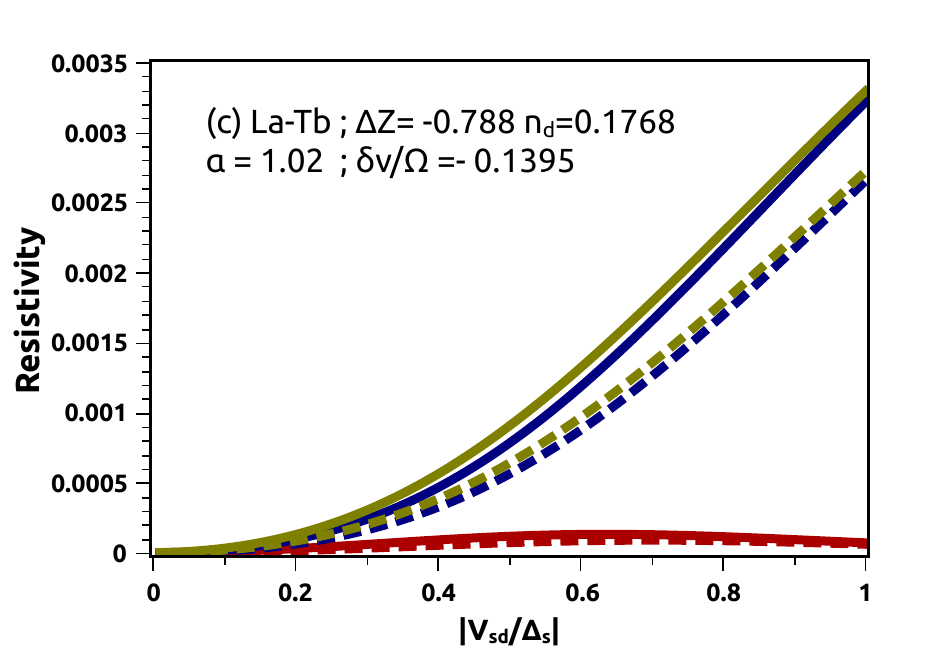} \includegraphics[scale=0.43]{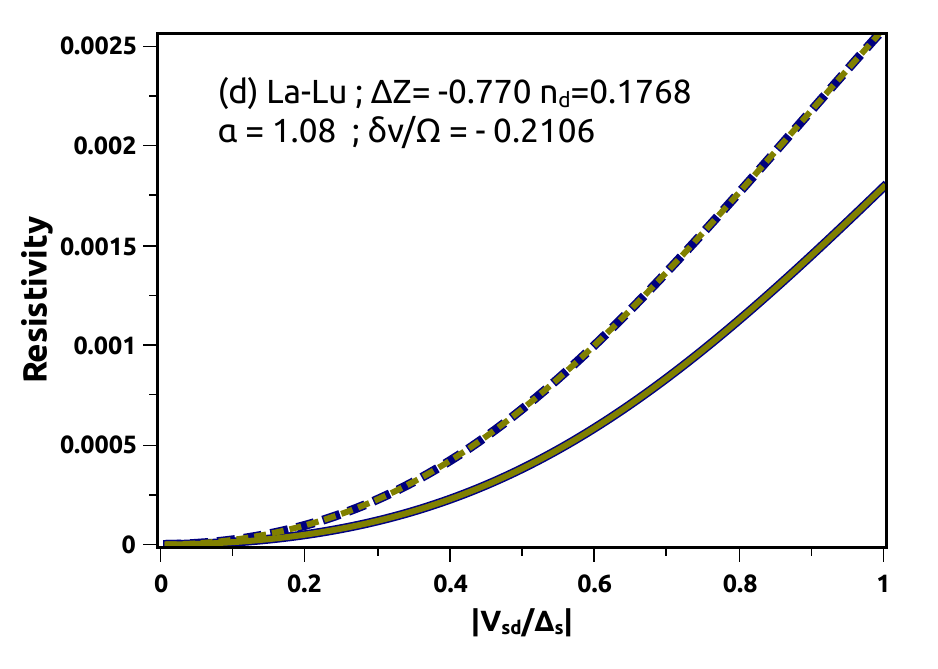}
\caption{Resistivity of La alloys \textit{vs} $\vert V_{sd}/\Delta _s\vert $ using the DFT band model. In (a) La-Gd, (b) La-Dy, (c) La-Tb and (d) La-Lu alloys.  With, the total resistivity in green, the residual resistivity in blue and the spin dependent resistivity in red lines, respectively.} 
\label{FIG14} 
\end{center}
\end{figure}

In figure \ref{FIG13}, we present the same results as in Figure \ref{FIG12} for our extended model; however, here, the results for all the studied alloys are displayed within the same panel, while each resistivity term is plotted in separate panels. In the figure (a) $R_0$ in equation (\ref{DELTAR0}) (blue lines), (b) $R_{DG-F}^{eff}$ in equation (\ref{JEFF}) (red lines) and (c) $R=R_0+R_{DG-F}^{eff}$ (green lines) using the parabolic band model. The solid, dotted, dashed and circles lines correspond to La-Tb, La-Gd, La-Dy and La-Lu respectively.

Considering the DFT band model, figure \ref{FIG14} summarizes the resistivity of the alloys in terms of $\vert V_{sd}/\Delta s\vert $ (analogous to figure \ref{FIG12}). In this figure, solid and dotted lines represent the cases $\alpha \neq 1$; $\delta v/ \Omega \neq 0$ and $\alpha = 1$; $\delta v/ \Omega = 0$, respectively. 
Similarly to the parabolic model, our DFT results for Gd in La, shown in Figure \ref{FIG14}, did not achieve convergence for Friedel's sum rule at $\alpha =1$ and $\delta v/ \Omega = 0$ across any of the parameter ranges explored in the DFT calculations performed using VASP. 

Notably, substantial differences were observed when compared to the ``parabolic'' model case in figure \ref{FIG12}. Our DFT results yield an effective De Gennes–Friedel resistivity, $R_{DG-F}^{eff}$, taking only positive values within the same $\vert V_{sd}\vert$ range as in the parabolic case. 
 
Another substantial difference with the parabolic case is that both, the total, $R$, and charge resistivity $R_0$ increase almost linearly with $V_{sd}$ (for $V_{sd}>0.6$), while in the parabolic case, the total and charge resistivity reach a plateau around $V_{sd}/\Delta _2 = 0.8$. 

Also, as shown in figures \ref{FIG14}(b) and (c), the influence of the spin term on the total resistivity is minor than in the parabolic case. Conversely, while the period and volume effects have a less noticeable impact on the spin resistivity, they still affect both the total and residual resistivities. It is important to note that unlike the parabolic case, the DFT calculations show that the period and volume effects decrease the value of the residual resistivity, consequently the total resistivity, for cases of La-Dy (b) and La-Lu (d).

\begin{figure} 
\begin{center}
\includegraphics[scale=0.46]{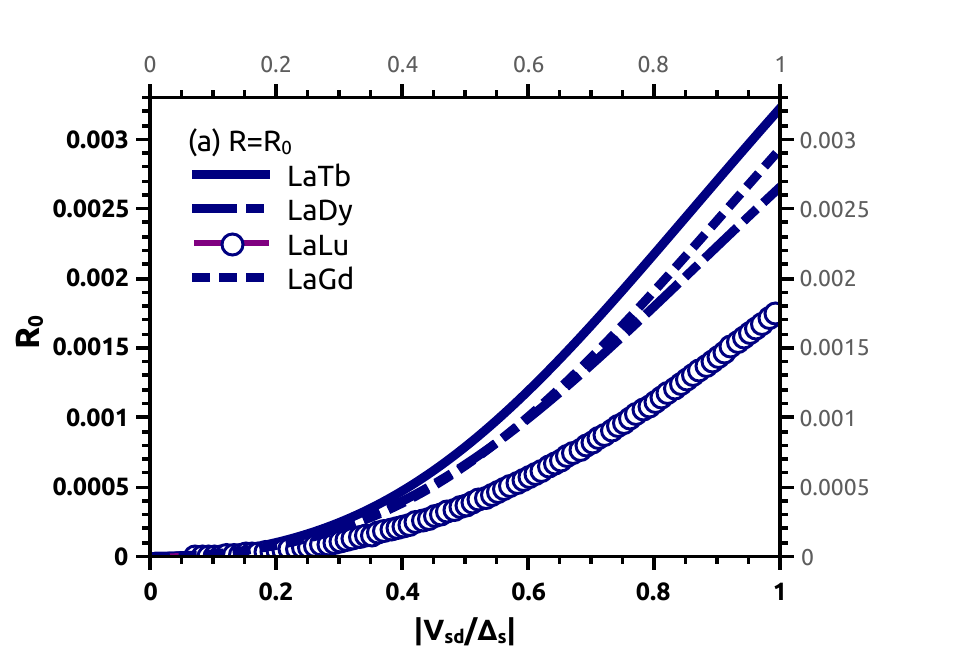} \includegraphics[scale=0.46]{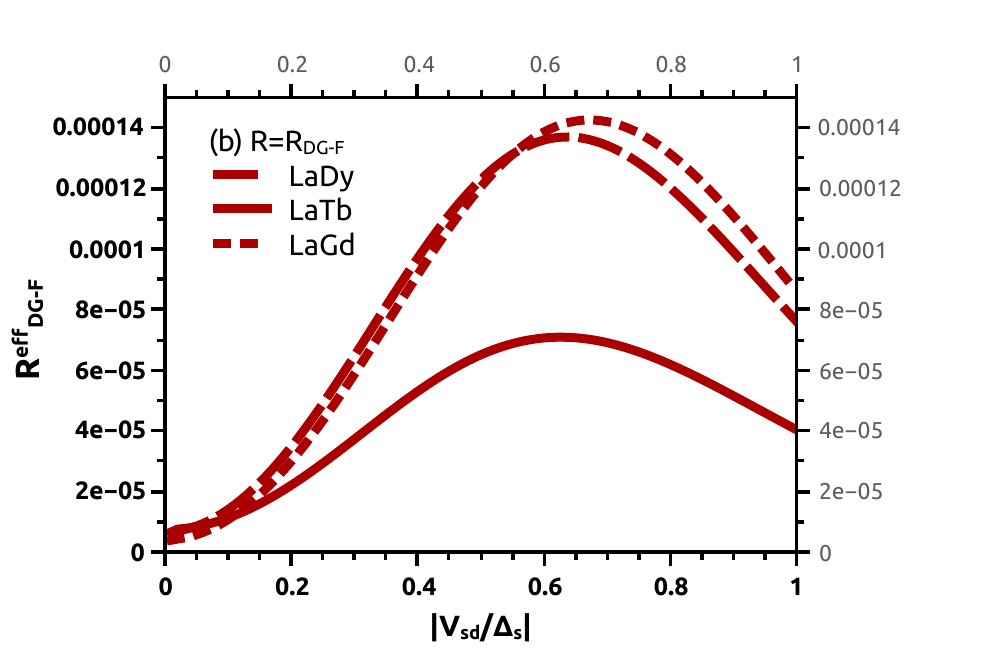} \\
\includegraphics[scale=0.46]{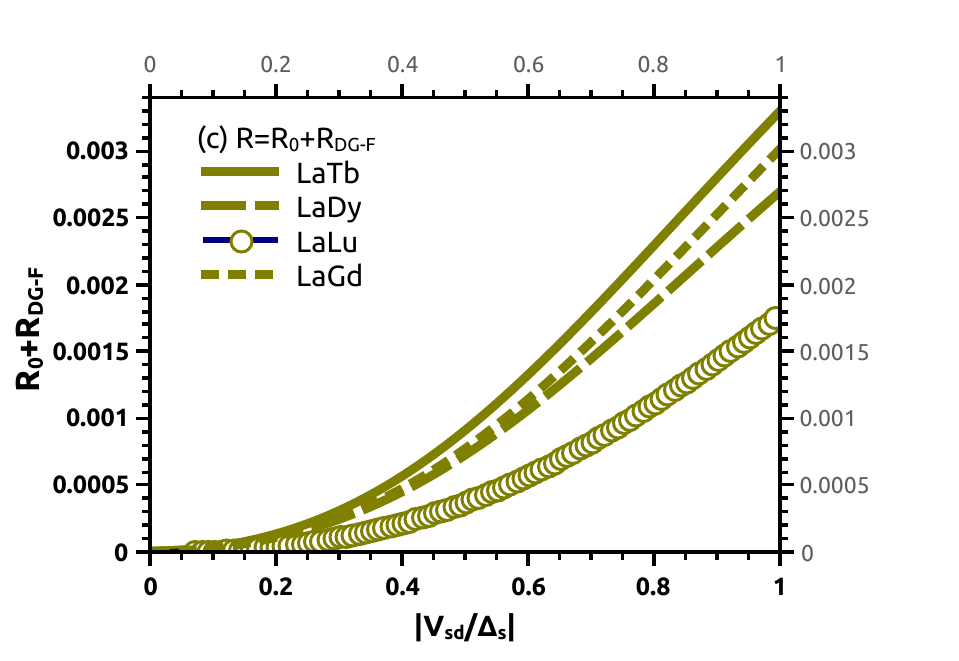} 
\caption{Comparison of the (a) residual resistivity $R_0$ (in blue), (b) the spin resistivity $R_{DG-F}^{eff}$ (in red), and (c) the total resistivity $R=R_0+R_{DG-F}^{eff}$ (in green), in terms of $\vert V_{sd}/\Delta _s\vert $ using the DFT band model. The solid, dotted, dashed and circles lines correspond to La-Tb, La-Gd, La-Dy and La-Lu respectively.} 
\label{FIG15} 
\end{center}
\end{figure}

In figure \ref{FIG15} (analogous to figure \ref{FIG13}), we present respectively each term contributing to the resistivity namely, (a) $R_0$ (blue) in equation (\ref{DELTAR0}), (b) $R_{DG-F}^{eff}$ (red) in equation (\ref{JEFF}) and (c) $R=R_0+R_{DG-F}^{eff}$ (green) separately using the band model from DFT calculations.
The solid, dotted, dashed and circles lines correspond to La-Tb, La-Gd, La-Dy and La-Lu respectively.

In summary, we have introduced a two-band model alongside two additional effects: the period effect and the volume difference introduced by the impurity in relation to the host atom. The inclusion of a two-band model gives rise to the  term $\theta (\epsilon _F,\vert V{sd}\vert ^2)$ in equation (\ref{HExacta}), which combines charge and spin effects. This combination not only induces a deviation in the standard De Gennes–Friedel resistivity term but also significantly modifies the total resistivity, as illustrated in Figures \ref{FIG12} and \ref{FIG14}. The effects are particularly evident in the case of Lu, which lacks a spin term, allowing for a clearer observation of the period and volume effects. This alloy thus highlights the importance of accounting for both period and volume effects in resistivity analysis.

\section{Comments and conclusion}
\label{S6}

Briefly, we have introduced a two-band model alongside two additional effects: the period effect and the volume difference introduced by the impurity in relation to the host atom. The inclusion of a two-band model gives rise to the  term $\theta (\epsilon _F,\vert V{sd}\vert ^2)$ in equation (\ref{HExacta}), which combines charge and spin effects. This combination not only induces a deviation in the standard De Gennes–Friedel resistivity term but also significantly modifies the total resistivity. The effects are particularly evident in the case of Lu, which lacks a spin term, allowing for a clearer observation of the period and volume effects. This alloy thus highlights the importance of accounting for both period and volume effects in resistivity analysis.

In addition, we have employed two different band structure models, namely: a ``parabolic'' band model and a more realistic one derived from first-principles calculations using the VASP code.

Within this approach, we have studied the influence of the band structure on the temperature-independent resistivity of  diluted lanthanum alloys containing rare earth impurities such as Gd, Tb, Dy, and Lu. 

Our results confirm a strong dependence of the resistivity with the band structure and, consequently, on the shape of the density of states (DOS) employed. We highlight several findings:

\begin{enumerate}[1.]

\item The calculated temperature-independent resistivity in terms of the hybridization charge potential, $\vert Vsd\vert ^2$, varies significantly when we use a parabolic band model or a more realistic one.
\item In the transition metals, the resistivity expression includes exchange parameters appearing as $(J^{(s)})^2$, $(J^{(d)})^2$ and $J^{(s)}\times J^{(d)}$. We have corroborated that these cross products significantly influence the effective exchange parameter $J^{(s)}_{eff}$ as defined in equation (\ref{JEFF}).
\item The presence of two bands (s and d) and two exchange couplings alters the conventional results of De Gennes and Friedel.
\item We have found substantial differences concerning DFT results compared to the parabolic model, where the behavior of resistivity drastically changes with the hybridization term $\vert V_{sd} \vert $.
\item The total resistivity, $R$, increase with $\vert V_{sd}\vert $ when we consider the realistic band whereas in the parabolic case both the total and charge resistivity reach a plateau for $\vert V_{sd} \vert > 0.8$.
\item The spin resistivity value varies drastically depending on the band model used. It shifts from behavior where resistivity can take both positive and negative values, whereas with the DFT model, resistivity only assumes positive values within the considered $\vert V_{sd} \vert$ range.
\end{enumerate}

From our model, it is evident that the introduction of a rare-earth impurity in the transition host exerts dual effects:

\begin{itemize}
\item The primary effect arises from the f level (assumed to be below the conduction s-d bands), which contributes a spin-dependent potential via exchange interactions $J^{(s)}$ and $J^{(d)}$.
\item The typical trivalent valence state of rare earths serves as a source of scattering, with particular emphasis on the host's nature and s-d bands existence.
\end{itemize}

Finally, given the significant impact of the DOS on resistivity, we will extend our analysis to Gd and Lu impurities diluted in transition metal hosts belonging to the 5d series, namely Hf, Ta, W, Re, Ir, Os, Pt, and Au. Concurrently, electronic transport calculations will be performed using VASP and the semi-classical Python module BoltzTrap2 \cite{BoltzTrap2}, which uses our band structure calculation results to determine the resistivity of these alloys. This approach will enable us to compare our numerical findings with those obtained with BoltzTrap2 when experimental data are unavailable. These calculations will be presented in an upcoming paper.

\section*{Acknowledgement}

One of the authors, V.P. Ramunni, would like to thank Camilo G. Rivas (a student in scientific computing) for his invaluable assistance in developing Python programs. 

\appendix
\section{Phase shift parameters}
\label{Ap1}
Taking into account the following definition
\begin{equation} 
 F_{\lambda }(\epsilon \pm i\delta )=F_{\lambda }^{R}(\epsilon)\mp F_{\lambda }^{I}(\epsilon), 
\label{F1}
\end{equation}
\noindent with $F_{\lambda }^{R}$ the real part of $F$
\begin{equation} 
F_{\lambda }^{R}(\epsilon)=P\sum _k \frac{1}{\epsilon-\epsilon _{k}^{(\lambda)} } = P^{\lambda }_{00}, 
\label{F2}
\end{equation}
\noindent and the imaginary part as
\begin{equation} 
F_{\lambda }^{I}(\epsilon)=\pi \rho _{\lambda }(\epsilon) ,
\label{F3}
\end{equation}
\noindent $\rho _{\lambda }(\epsilon)$ represents the density of states of the conduction electrons. Introducing the expressions for the phase shift, we can write:
\begin{equation} 
F_{\lambda }(\epsilon \pm i\delta )=\vert   F_{\lambda }(\epsilon)\vert \exp \left\{\mp i\delta (\epsilon)\right\},
\label{F4} 
\end{equation}
\noindent where
\begin{equation} 
\vert   F_{\lambda }(\epsilon)\vert   =\left\{ [F_{\lambda }^{R}(\epsilon)]^2+[F_{\lambda }^{I}(\epsilon)]^2 \right\}^{1/2} ,
\label{F5}
\end{equation}
\noindent with
\begin{equation} 
\cos \delta _{\lambda } (\epsilon) = \frac{F_{\lambda }^{R}(\epsilon)}{\vert   F_{\lambda }(\epsilon)\vert   },
\label{F6} 
\end{equation}
\noindent and
\begin{equation} 
\mbox{sen} \delta _{\lambda } (\epsilon) = \frac{F_{\lambda }^{I}(\epsilon)}{\vert   F_{\lambda }(\epsilon)\vert   }. 
\label{F7}
\end{equation}
\begin{equation}        
F_s (\epsilon \pm i\delta _s )F_d (\epsilon \pm i\delta _d )=\vert   F_s (\epsilon)\vert   \vert   F_d (\epsilon)\vert   \exp \left\{\mp [\delta _s+\delta _d]\right\} . 
\label{F8}
\end{equation}
\noindent Similarly, 
\begin{equation} 
1-V_{dd}(\epsilon) F_{d}(\epsilon)=\vert   1-V_{dd}(\epsilon)F_{d}(\epsilon)\vert \exp \left\{\mp i\eta _{dd}(\epsilon)\right\}, 
\label{F9}
\end{equation} 
\noindent where
\begin{equation} 
\vert   K(\epsilon)\vert   =1-V_{dd}(\epsilon)F_d(\epsilon)=\left\{ 
[1-V_{dd}(\epsilon)F_d ^R(\epsilon)]^2+[V_{dd}(\epsilon)F_d ^I (\epsilon)]^2 \right\}^{1/2}. 
\label{F10}
\end{equation} 
\noindent Again, we define the functions:
\begin{eqnarray} 
\cos \left[\eta _{dd}(\epsilon)\right] & = & \frac{1-V_{dd}(\epsilon)F_d^{R} (\epsilon) }{\vert   K(\epsilon)\vert   }, \label{F11} \\ 
\mbox{sen} \left[\eta _{dd}(\epsilon)\right] & = & -\frac{V_{dd}(\epsilon)F_d^{I} (\epsilon)}{\vert   K(\epsilon)\vert   } \label{F12}
\end{eqnarray} 
\noindent In the following section, we present some expressions derived in Ref. \cite{Atroper} to illustrate how these expressions are modified under the approximations explored in the present work.

\begin{equation}
1-V_{dd}(\epsilon)F_{d}(\epsilon)-\vert   V_{sd}\vert   ^2 F_s (\epsilon)F_d (\epsilon)=\vert   K(\epsilon)\vert   e^{\mp i\eta (\epsilon)},
\label{F13} 
\end{equation}
\noindent where
\begin{eqnarray} 
\vert   K(\epsilon)\vert    & = & \left\{ \left[1-V_{dd}(\epsilon)F_{d}^ R (\epsilon)-\vert   V_{sd}\vert   ^2 
[F_d ^R (\epsilon)F_s ^R (\epsilon)-F_d ^I (\epsilon)F_s ^I (\epsilon) ] \right] ^2 \right. \nonumber \\ 
& + & \left. \left[-V_{dd}(\epsilon)F_{d}^I (\epsilon)-\vert   V_{sd}\vert   ^2 [F_d ^R 
(\epsilon)F_s ^I (\epsilon)+F_d ^I (\epsilon)F_s ^R (\epsilon)] \right] ^2 \right\}^{1/2} .
\label{F14}
\end{eqnarray} 
\noindent Then, 
\begin{equation} 
\cos \left[\eta (\epsilon)\right] = \frac{\left[1-V_{dd}(\epsilon)F_{d}^ R (\epsilon)-\vert   V_{sd}\vert   ^2 
[F_d ^R (\epsilon)F_s ^R (\epsilon)-F_d ^I (\epsilon)F_s ^I (\epsilon)] \right]}{\vert   K(\epsilon)\vert   }
\label{F15}
\end{equation}
\noindent and
\begin{equation}
\sin \left[\eta (\epsilon)\right] = -\frac{\left[V_{dd}(\epsilon)F_{d}^I (\epsilon)-\vert   V_{sd}\vert   ^2 
[F_d ^R (\epsilon)F_s ^I (\epsilon)+F_d ^I (\epsilon)F_s ^R (\epsilon)] \right]}{\vert   K(\epsilon)\vert   }.
\label{F16} 
\end{equation}

\section{Extended Friedel's sum rule}
\label{Ap2}

The total change in the density of states resulting from the introduction of
 the impurity can be calculated as the difference between the imaginary part of the perturbed Green function, $G_{jj}^\lambda (\epsilon)$, and the Cauchy's principal part $P_{jj}^\lambda (\epsilon)$  defined in equation (\ref{F2}), summed over all sites.
The Green function, 

\begin{eqnarray}  
G_{jl}^\lambda (\epsilon) &=&P_{jl}^\lambda (\epsilon)+P_{j0}^\lambda (\epsilon)\frac{V^{\lambda}(\epsilon)}{\alpha  
^2-V^{\lambda}(\epsilon)P_{00}^\lambda (\epsilon)}P_{0l}^\lambda (\epsilon)+  \nonumber  
\\  
&&-\left( \alpha -1\right) \frac{\left( \alpha -1\right) P_{00}^\lambda  
(\epsilon)\delta _{j0}\delta _{0l}-\alpha \left( \delta _{j0}P_{0l}^\lambda  
(\epsilon)+P_{j0}^\lambda (\epsilon)\delta _{0l}\right) }{\alpha ^2-V^{\lambda}(\epsilon)P_{00}^\lambda (\epsilon) 
},  \label{eq:dyson4}  
\end{eqnarray}  
describes the electron's jump from site $j$ to site $l$, scattered by an effective charge potential $V^\lambda (\epsilon)$, at the impurity site,  
\begin{equation}  
V^\lambda (\epsilon)=\left( \alpha ^2-1\right) \left( \epsilon-\varepsilon _h\right)  
+V_0^\lambda \quad .  \label{eq:delz2}
\end{equation}  
\noindent After some algebraic manipulation, the propagator can be expressed as follows:

\begin{eqnarray}  
\Delta \rho ^{\lambda }(\epsilon) &=&-\frac 1\pi \mbox{Im}\sum_j\left( G_{jj}^{\lambda  }
(\epsilon)-P_{jj}^\lambda (\epsilon)\right)   \label{eq:2delrho} \\  
&=&-\frac 1\pi \mbox{Im}\left\{ \frac{-\left( \alpha ^2-1\right)  
P_{00}^\lambda (\epsilon)+V^{\lambda}(\epsilon)\sum_jP_{j0}^\lambda (\epsilon)P_{0j}^\lambda (\epsilon)}{%
\alpha ^2-V^{\lambda}(\epsilon)P_{00}^\lambda (\epsilon) }\right\}.  \nonumber  
\end{eqnarray}  
Remembering that,  
\begin{equation}  
\sum_jP_{j0}^\lambda (\epsilon)P_{0j}^\lambda (\epsilon)=-\frac \partial {\partial  
\epsilon}P_{00}^\lambda (\epsilon).  
\label{eq:gi0}  
\end{equation}  
By replacing equation (\ref{eq:gi0}) into (\ref{eq:2delrho}), we obtain the total change in the density of states of the perturbed system as a function of energy:
\begin{equation}  
\Delta \rho ^\lambda (\epsilon)=-\frac 1\pi \mbox{Im}\left\{ \frac{-\left( \alpha  
^2-1\right) P_{00}^\lambda (\epsilon)-V^\lambda (\epsilon)\partial P_{00}^\lambda  
(\epsilon)/\partial \epsilon}{\alpha ^2-V^\lambda (\epsilon)P_{00}^\lambda (\epsilon) }\right\}.  
\label{eq:deltarho2}  
\end{equation}  

The variation in the electron occupation number with spin $\lambda $, denoted as $\Delta N^\lambda $, is obtained through,
\begin{equation}  
\Delta N^\lambda =\int_{-\infty }^{\varepsilon _F}\Delta \rho ^\lambda  
(\epsilon)d\epsilon=-\frac 1\pi \mbox{Im}\ln \left\{ \alpha ^2-V^\lambda (\varepsilon _F)P_{00}^\lambda (\varepsilon _F)\right\}. \label{difqzeller1}
\end{equation}  
Evaluated at the Fermi level $\epsilon = \epsilon _F$, the $V_0^\lambda $ potentials are calculated self-consistently from the shielding total charge difference as follows, 
\begin{equation}  
\Delta Z=\Delta N^{\uparrow }+\Delta N^{\downarrow }, \label{difqzeller2}
\end{equation}  
with $\Delta N^{\lambda }$ from equation (\ref{difqzeller1}).

Note that, in equation (\ref{difqzeller2}), $\Delta Z$ is fixed, while $\Delta N^\lambda $ in equation (\ref{difqzeller1}) is a variable that depends on the value of $V_0^\lambda $, which is calculated self-consistently. The self-consistency is achieved as follows: we start with an initial of $V_0^\lambda$ that satisfies the Daniel-Friedel condition; the corresponding value of $\Delta N$ is then obtained for these potentials using equation (\ref{eq:delz2}). This value must equal the charge difference $\Delta Z$ introduced by the impurity.  If it does not, we select new values for the potentials $V_0^\sigma $ and repeat the procedure until the condition $\Delta N=\Delta Z$ is satisfied.

\bibliographystyle{elsarticle-num}

\end{document}


Additionally, to DFT calculations for the primitive dhcp La cointaining four atoms, we provide the density of states and band structure for the lanthanum supercell $La_{16}$, which contains 16 atoms. The results of the state densities and band structures of the alloys will not be used in the present work but will be relevant when we study electronic transport with BoltzTrap2 in a future work.

\begin{figure}
\begin{center}
\includegraphics[scale=0.6]{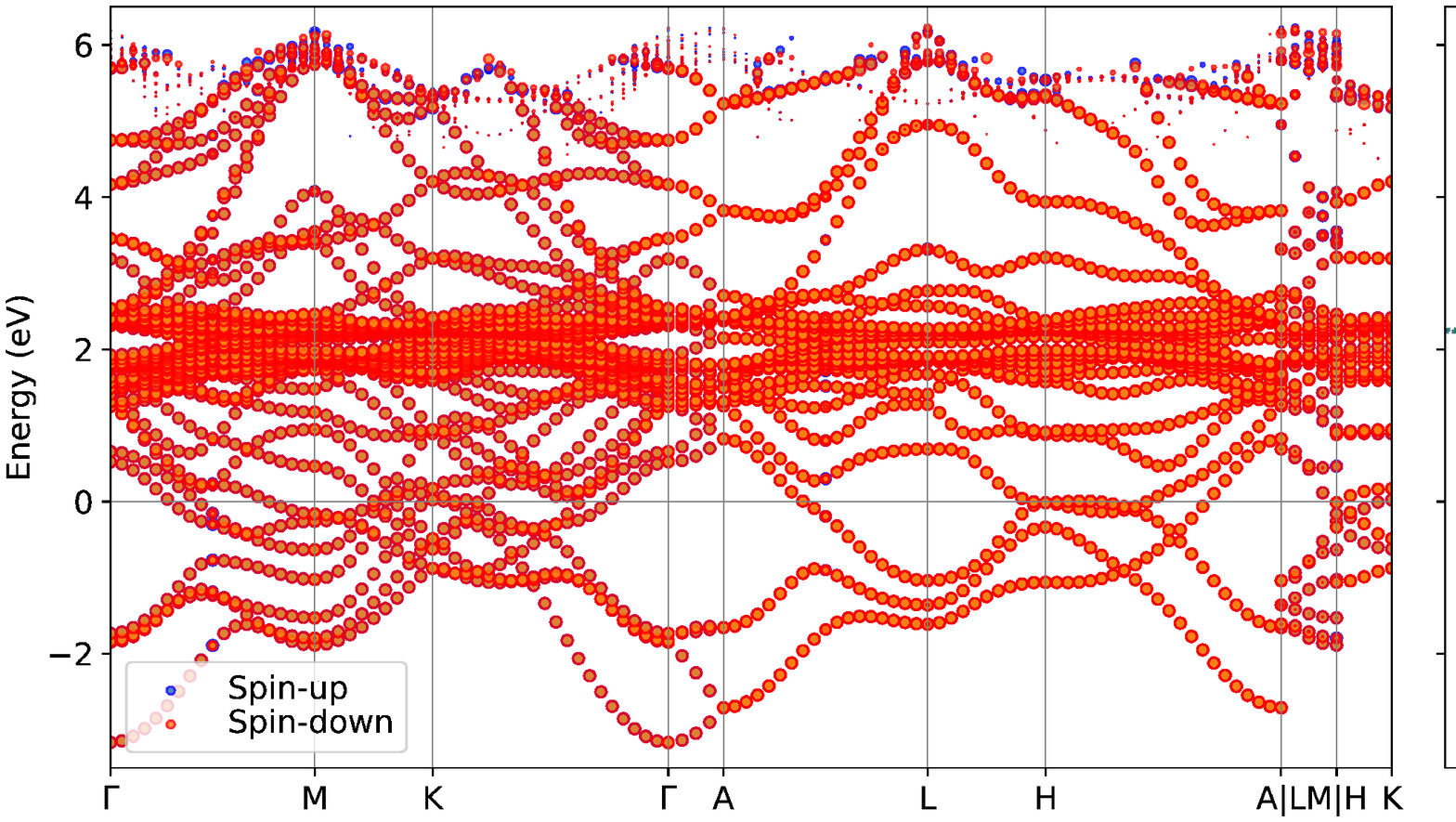}
\caption{Density of states and band structure for the Lanthanum supercell containing 16 atoms.} 
\label{FIG5} 
\end{center}
\end{figure}

Figure \ref{FIG5} displays the Density of States (DOS) and Band Structure (BS) for the Lanthanum Supercell Containing 16 atoms. The left panel illustrates the band structure, while the right panel presents the density of states. We constructed the alloys here studied from this supercell, consisting of 15 La atoms and one impurity atom positioned at a substitutional site within the $La_{16}$ supercell. These results provide insights into the electronic properties of the supercell, which serves as the basis for studying the effects of impurities in the alloys.

We obtain the density of states (DOS) for diluted dhcp-La alloys, specifically La${15}$Gd, La${15}$Dy, La${15}$Tb, and La${15}$Lu. The calculations were performed for a supercell of 16 atoms, constructed from the coordinates of the primitive cell of La$_4$, using Vaspkit~\cite{VASPKIT} with parameters $n_x = n_y = 2$ and $n_z = 1$. In these alloys, the impurities are incorporated at substitutional sites. 

For the DOS and band structure calculations, we first relaxed the supercell containing one impurity atom (La$_{15}$I), focusing only on the atomic positions while keeping the supercell's size and volume fixed. To enhance the accuracy of the DOS calculations, we doubled the number of k-points compared to the bulk relaxation. The high-symmetry k-points for the alloy were obtained for UNFOLDING calculations using the Vaspkit utility, and we later verified them using the SEEK site~\cite{SEEK}.

\begin{figure}
\begin{center}
\includegraphics[scale=0.4]{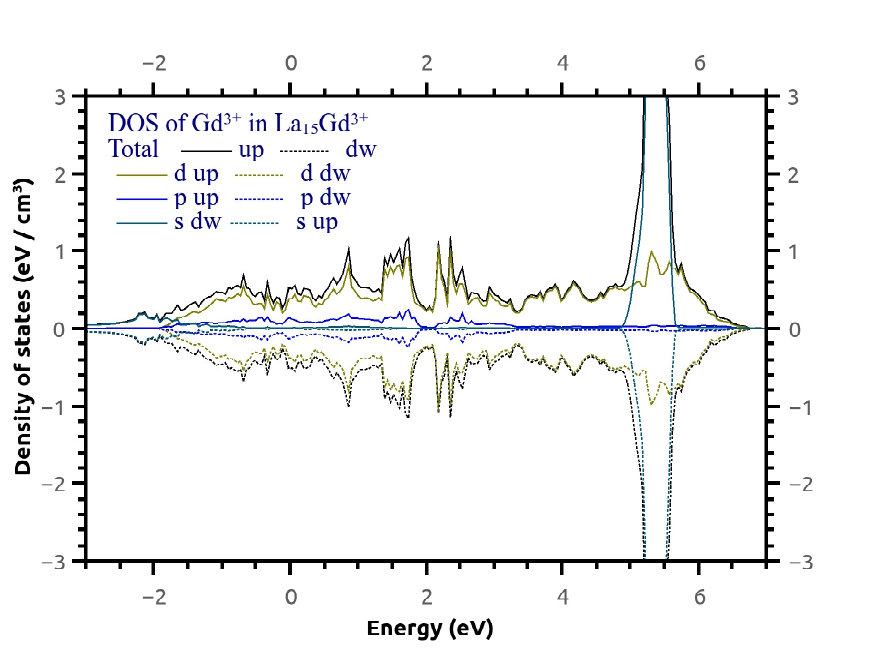}
\includegraphics[scale=0.4]{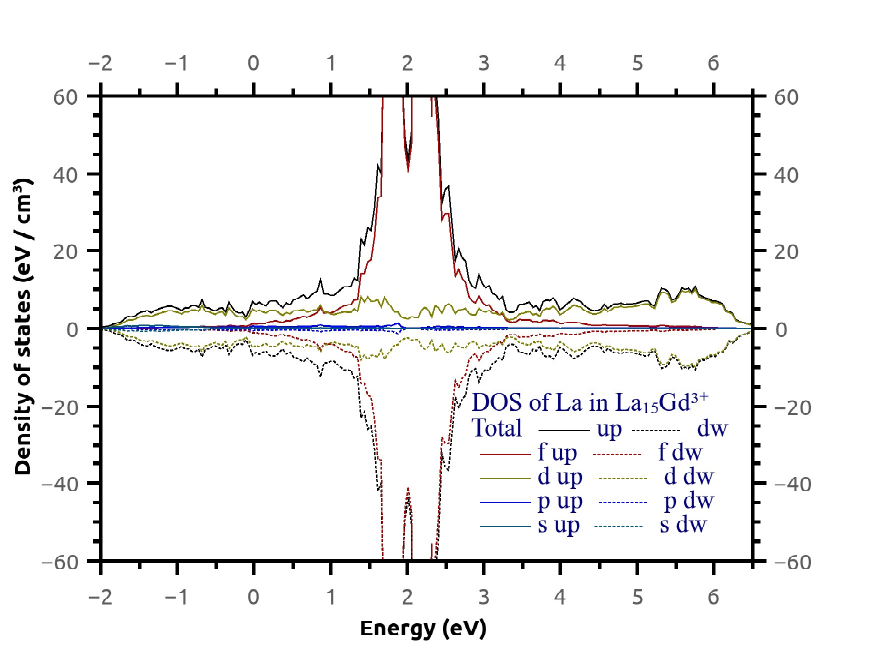} \\
\includegraphics[scale=0.4]{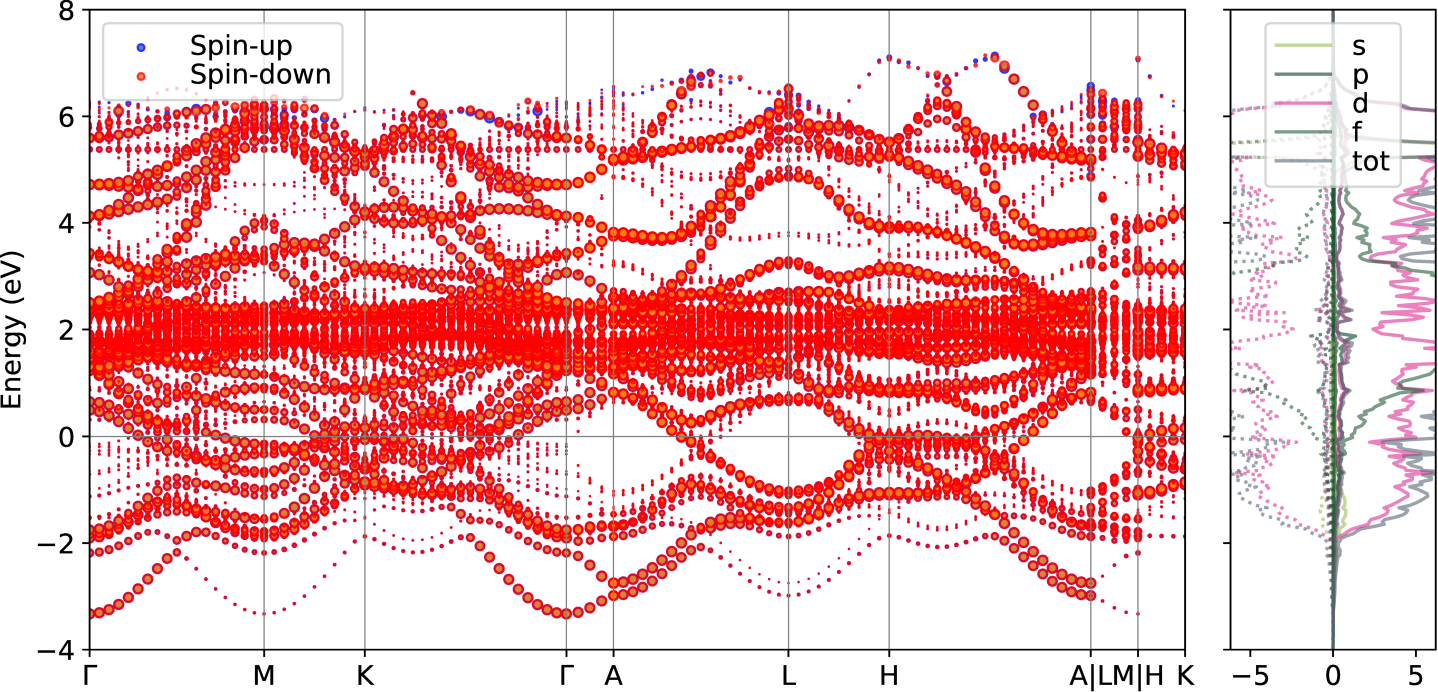}
\caption{Density of states (in eV/cm$^3$) \textit{vs} Energy (in eV) for La$_{15}$Gd: The plots display the s-p bands (Left) and the d-band (Right) for both La and Gd.} 
\label{FIG8}
\end{center}
\end{figure}

\begin{figure}
\begin{center}
\includegraphics[scale=0.4]{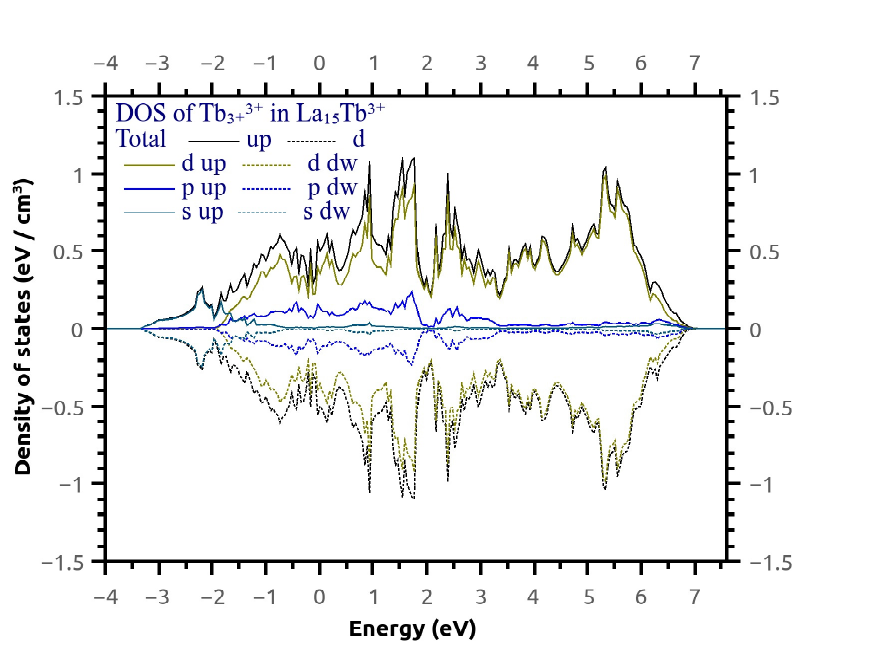}
\includegraphics[scale=0.4]{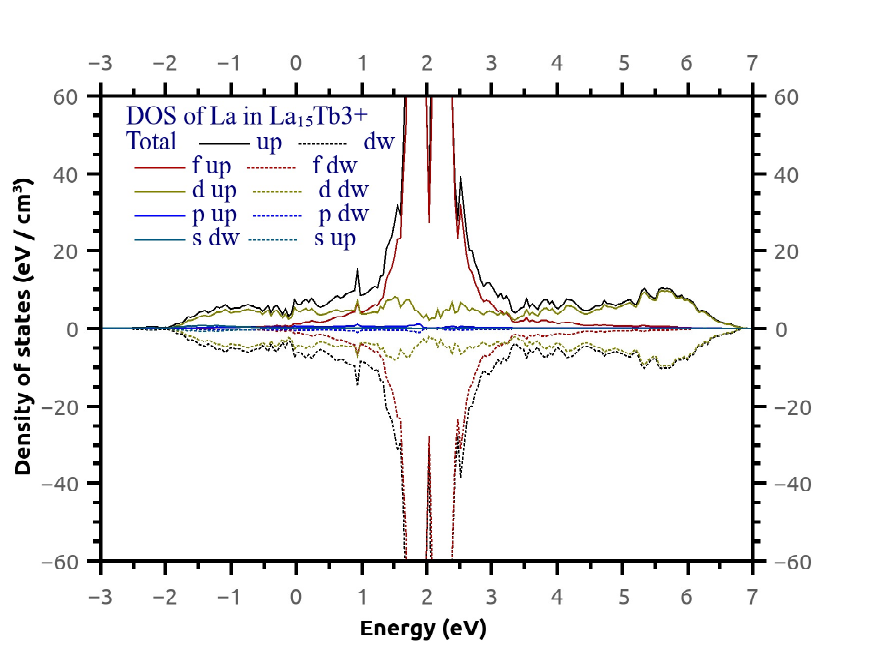} \\
\includegraphics[scale=0.4]{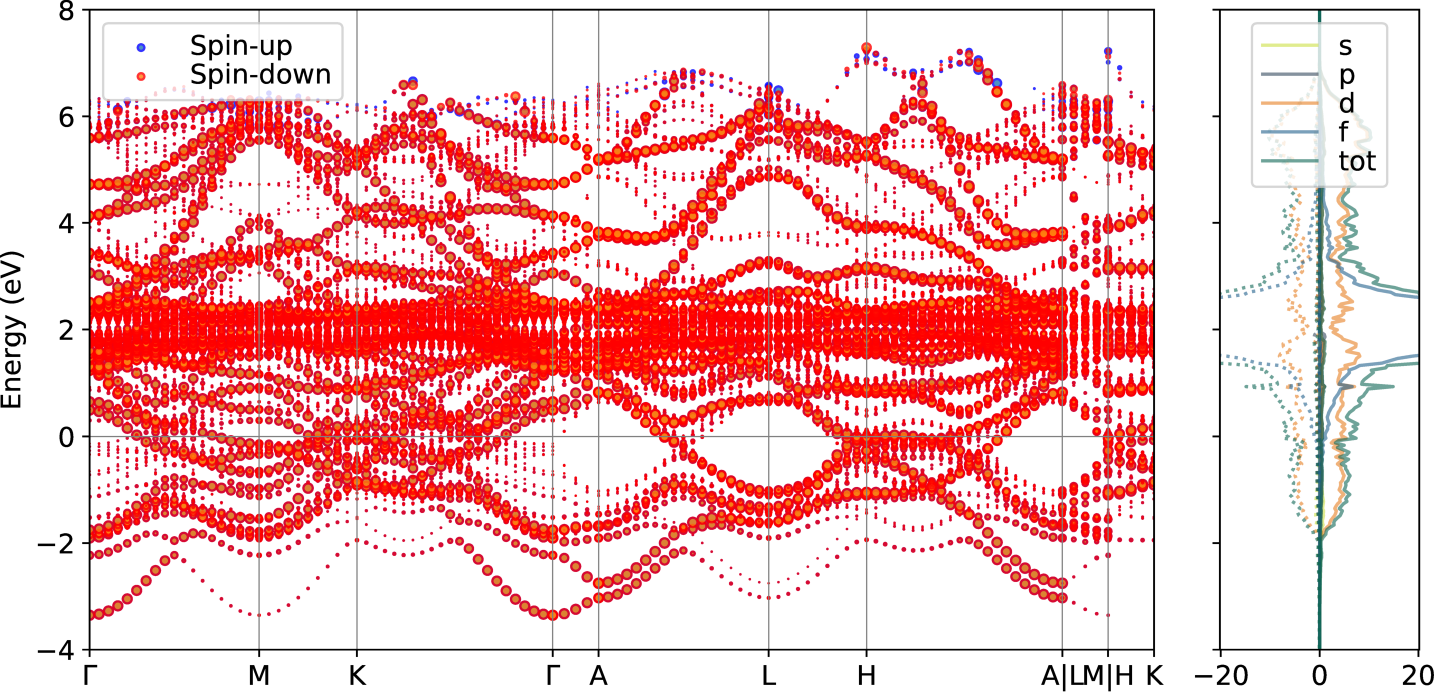}
\caption{Density of States (in eV/cm$^3$) vs. Energy (in eV) for La$_{15}$Tb: The plots illustrate the s-p bands (Left) and the d-band (Right) for both La and Tb.} 
\label{FIG9} 
\end{center}
\end{figure}

\begin{figure}
\begin{center}
\includegraphics[scale=0.5]{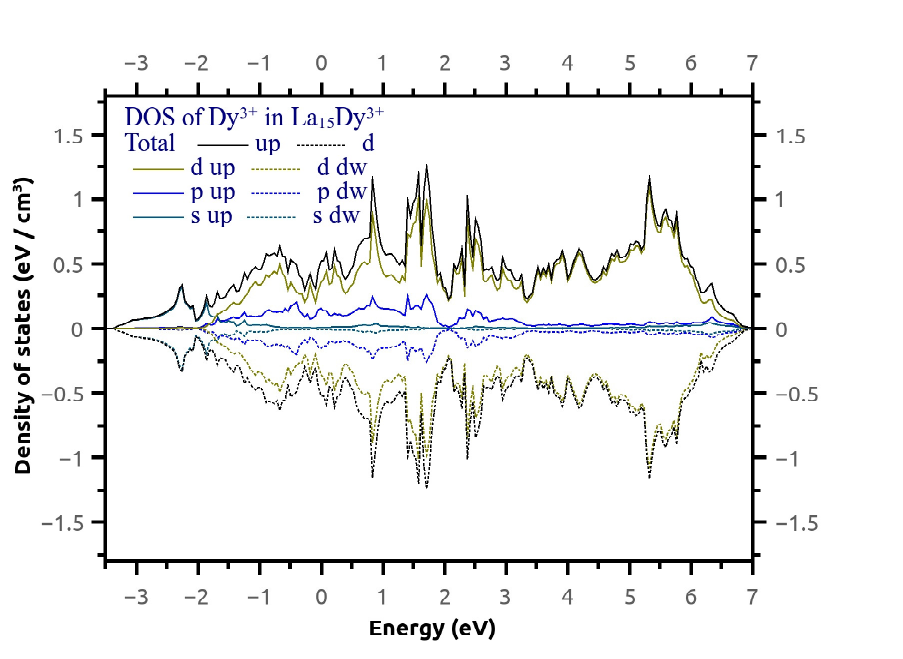}
\includegraphics[scale=0.5]{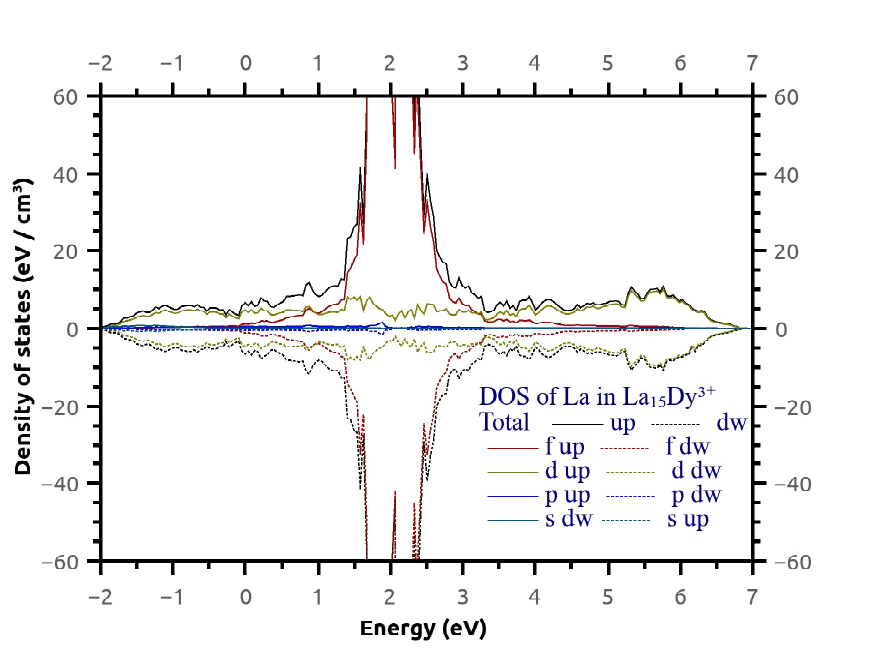} \\
\includegraphics[scale=0.5]{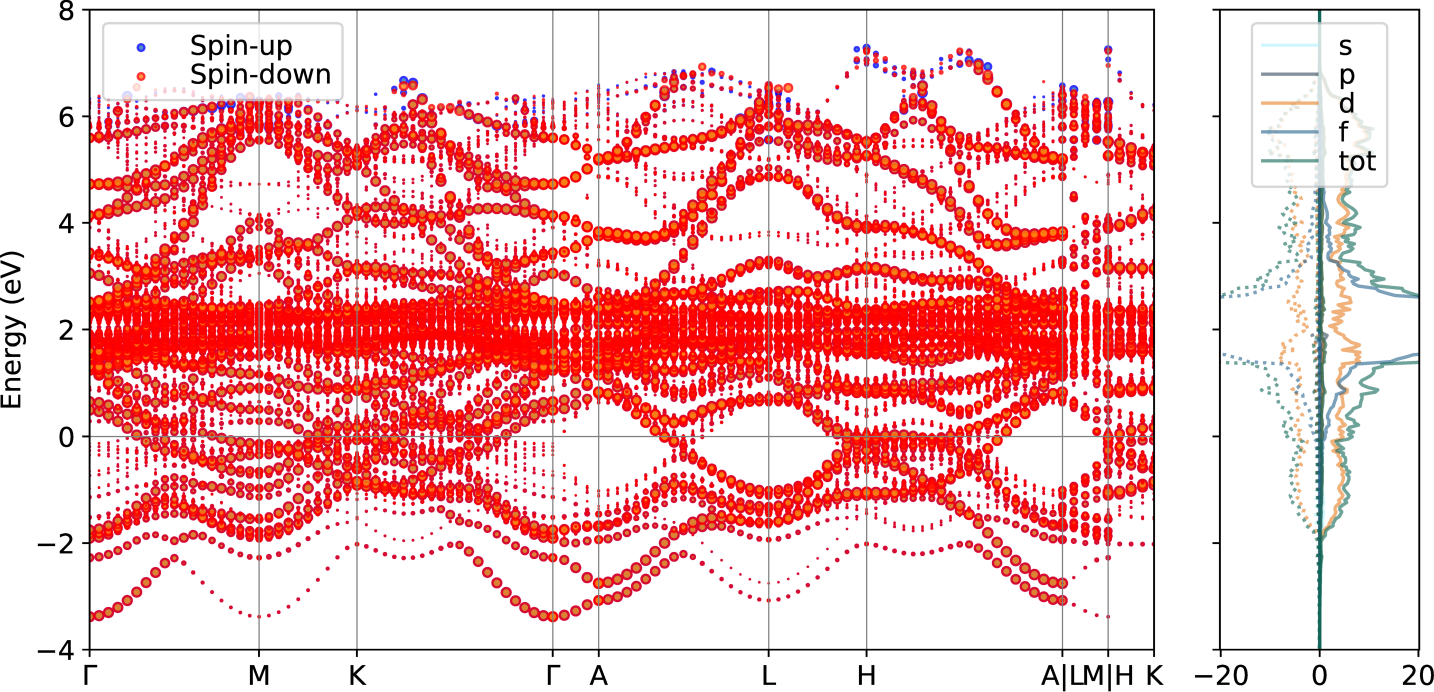}
\caption{Density of States (in eV/cm$^3$) vs. Energy (in eV) for La$_{15}$Dy: The plots illustrate the s-p bands (Left) and the d-band (Right) for both La and Dy.} 
\label{FIG10} 
\end{center}
\end{figure}

\begin{figure}
\begin{center}
\includegraphics[scale=0.5]{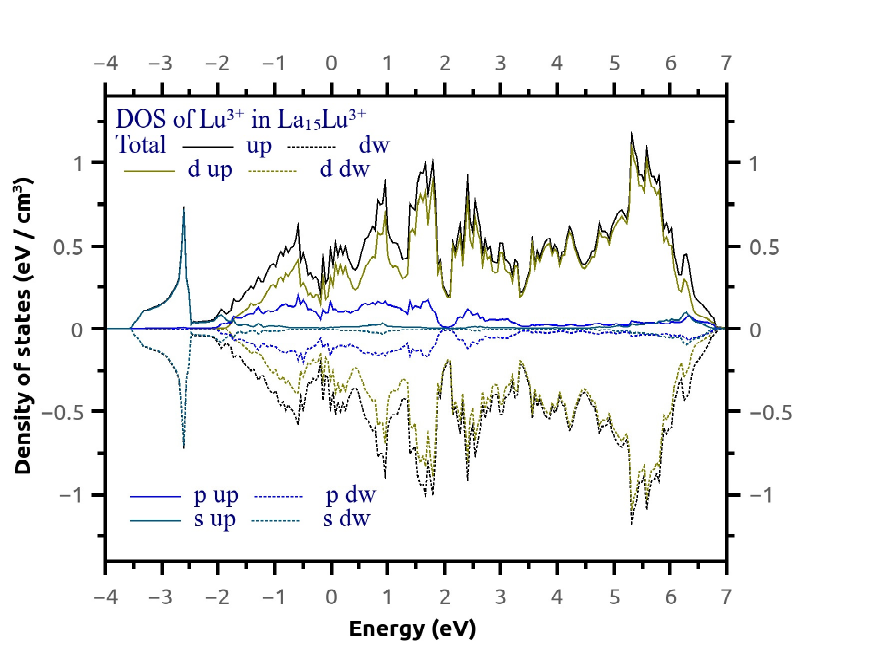}
\includegraphics[scale=0.5]{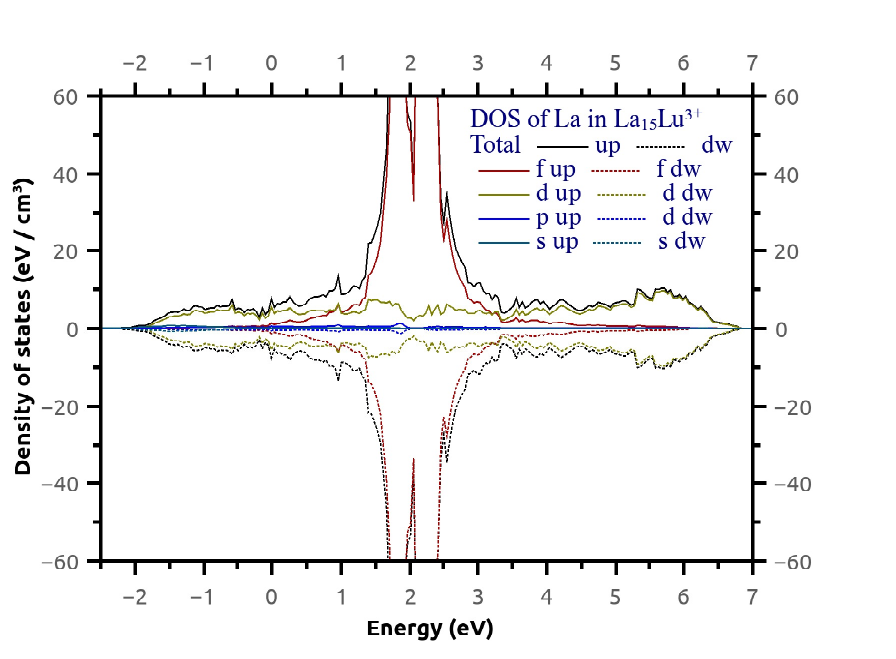} \\
\includegraphics[scale=0.5]{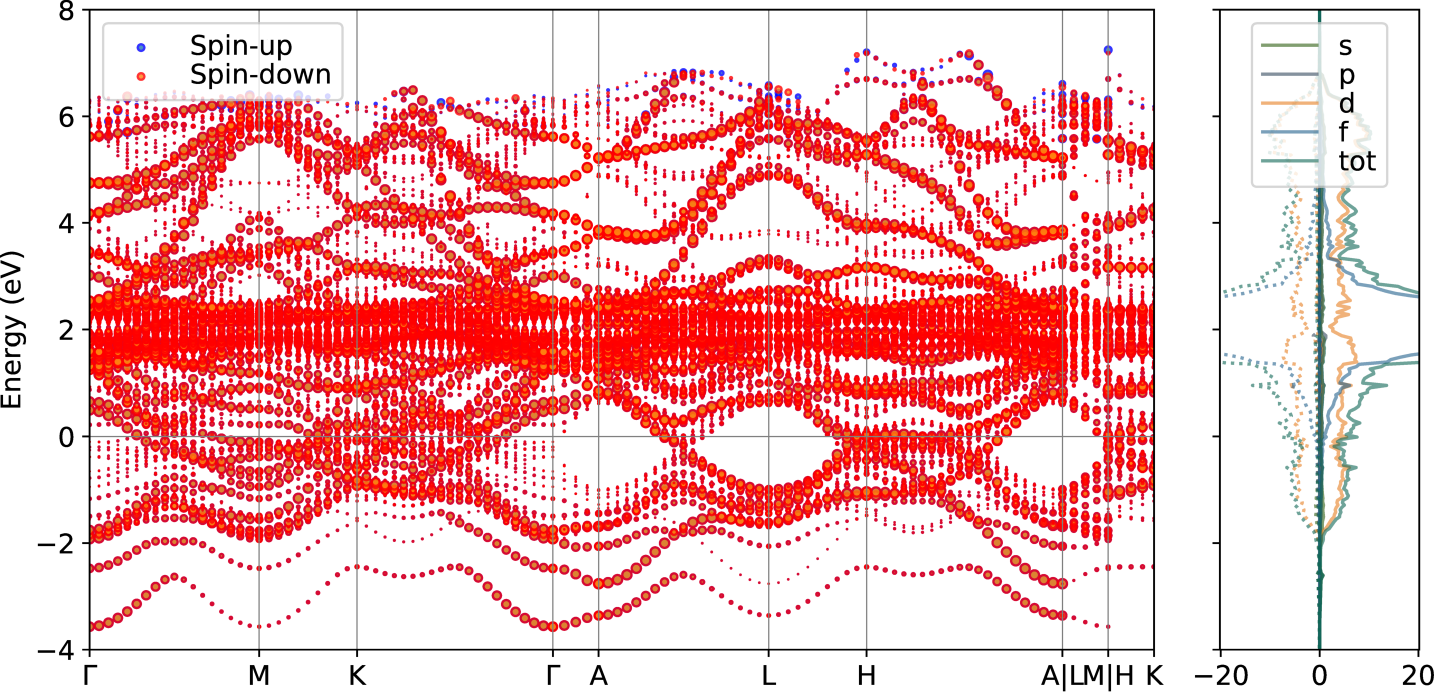}
\caption{Density of States (in eV/cm$^3$) vs. Energy (in eV) for La$_{15}$Lu: The plots illustrate the s-p bands (Left) and the d-band (Right) for both La and Lu.} 
\label{FIG11} 
\end{center}
\end{figure}

In figures \ref{FIG8} to \ref{FIG11}, we present our DFT results for the density of states and band structures of the dilute La alloys containing substitutional rare earth impurities: Gd, Tb, Dy, and Lu. These figures represent the four alloys studied in this work. Although the density of states (DOS) and band structures (BS) of the alloys are not utilized in our analysis, we include these results here to identify any significant differences compared to pure Lanthanum. In future work, we plan to calculate the electrical conductivity of the systems studied here using the Python module BoltzTrap2, which uses our present band structure calculations presented in these figures. The BoltzTrap2 module estimates conductivity based on the semi-classical Boltzmann theory of electronic transport \cite{BoltzTrap2}. This approach will enable us to check our theoretical predictions when no experimental data is available.

\bibliographystyle{elsarticle-num}